# A numerical modeling of wave-inclined slats interaction for particle methods

Lucas Soares Pereira[a], Rubens Augusto Amaro Jr.[b], Liang-Yee Cheng[a], Fabricio Simeoni de Sousa[b], Gustavo Massaki Karuka[c]

[a]Department of Construction Engineering, Polytechnic School at the University of São Paulo, Av. Prof. Almeida Prado, trav. 2, 83 – Cidade Universitária, São Paulo, 05508-900, SP, Brazil

[b]Department of Applied Mathematics and Statistics, Institute of Mathematical and Computer Sciences, University of São Paulo, Av. Trab. São Carlense, 400 – Centro, São Carlos, 13566-590, SP, Brazil

[c]MODEC, Inc., Nihonbashi Maruzen Tokyu Building 4th & 5th Floors 3-10 Nihonbashi 2-chome, Chuoku, Tokyo, 103-0027, Japan

lucas_pereira@usp.br, rubens.amaro@usp.br, cheng.yee@usp.br, fsimeoni@icmc.usp.br, gustavo.karuka@modec.com

**Abstract**

MPS-VG, a Virtual Grating (VG) model for the Lagrangian mesh-free Moving Particle Semi-implicit (MPS) method is proposed for replacing conventional particle-based solid modeling of gratings with a set of thin inclined slats. Unlike most approaches for perforated wave energy dampening devices, in which the flow through the device is simplified by pressure loss or damping effects without flow deflection, MPS-VG models the angular deviation caused by hydrodynamic impact on inclined slats. Both accuracy and computational performance of the model were checked through a simulation of wet dam break scenarios with the grating structures placed horizontally or vertically. The results were compared with those from fully particle-based modeling. MPS-VG correctly predicted complex wave-structure interactions using a relatively low-resolution model and significantly reduced processing time and memory storage compared to conventional particle-based MPS modeling. The evaluation of the performance of the gratings with inclined slats as wave energy dampers revealed the horizontal gratings outperformed the vertical ones. Therefore, qualitative and quantitative agreements strengthened the potential of MPS-VG as a practical and computationally efficient tool for the study of multi-scale phenomena of wave impacts on grating with inclined slats.

*Keywords: Breaking wave impact, grating, MPS, SPH, particle methods*

## 1. Introduction

Coastal and offshore structures such as ports, ships, oil and gas platforms, and marine renewable energy systems are naturally exposed to wave impact loads. Such forces can pose significant threats, potentially leading to catastrophic damages or stability issues (Temarel, et al., 2016). Among the devices that mitigate wave-induced loads, perforated plates (Cho & Kim, 2008) and flat bar or slat screens (Weber, et al., 2000), including V-shape, vane-type (Buchner, 2002), and perforated breakwaters (Amaro Jr, et al., 2019) or sloshing dampers (Bellezi, et al., 2019) are lightweight and cost-effective solutions that enable partial water circulation.

However, the hydrodynamics of wave energy dampers can be affected by small features such as tiny openings or an array of slats, posing a challenge for evaluations of their effectiveness. Regarding slats, flow deflection may occur, depending on the angle of attack. Therefore, global and local phenomena with distinct spatial scales should be considered for a realistic analysis of the complex Wave-Structure Interactions (WSI). This study proposes an efficient numerical modeling of the wave impact on inclined slats considering flow deflection.

Traditionally, experimental investigations and field measurements provide relevant data for WSI engineering applications. However, the reduced scale modeling which encompasses different spatial length scales is challenging regarding the dealing with slatted grating structures. Furthermore, concerns over scale effects related to small free surface piercing elements subjected to impulsive loads have arisen. Due to remarkable advances in high-performance computing systems, Computational Fluid Dynamics (CFD) has become an attractive complementary or alternative tool to address such problems (Huang, et al., 2022).

Mesh-free particle-based methods, such as Smoothed Particle Hydrodynamics (SPH) (Gingold & Monagham, 1977; Lucy, 1977) and Moving Particle Semi-implicit (MPS) (Koshizuka & Oka, 1996) have been widely adopted for WSI simulations owing to their ability in dealing with large free surface deformations and complex fluid-solid interfaces, as reported by Mazhar et al. (2021). Most particle methods employ a single uniform resolution across the entire computational domain. Therefore, high-resolution models are required to model multi-scale WSI problems that involve wave energy dampers with tiny local features. Significant efforts have been devoted to computationally efficient solutions towards avoiding high computational costs or unfeasible simulations.

Multi-resolution approach, a strategy that has drawn special attention, models the computational domain near tiny features using high resolution (small particles), while coarse resolution (large particles) is applied to the remaining domain. Several multi-resolution techniques have been proposed for mesh-free particle-based methods - some in the context of weakly-compressible formulations (Barcarolo, et al., 2014; Gao, et al., 2023; Sun, et al., 2023) and others considering incompressible (projection-based) approaches (Shibata, et al., 2017; Tanaka, et al., 2018; Bellezi, et al., 2022). However, computational challenges have arisen due to the complexity of multi-resolution algorithms and the accuracy, robustness, and computational efficiency of particle-based multi-resolution techniques remain topics of intense research.

On the other hand, a simple and computationally cost-saving approach is the inclusion of additional terms on fluid motion, such as pressure drop coefficients (Francis, et al., 2023; Faltinsen, et al., 2010), Morison's equation (McNamara, et al., 2021), and Ergun's equation (Awad & Tait, 2022), which is particularly well-suited for the prediction of interactions of waves with thin perforated walls, slat screens, or porous media. Nonetheless, a limitation of existing solutions is the flow deflection caused by wave-dampening devices is not considered. As a result, wave impact on grating structures with an array of inclined slats, e.g., vane-type breakwaters, bulwark-type structures, and ventilation panels, which are used on riser balconies or on decks of floating structures, cannot be correctly represented.

The present study has been motivated by the challenges for simulating the multi-scale WSI phenomenon involving lightweight wave dampers formed by grating structures with thin inclined slats that deflect wave flow. A novel modeling technique called Virtual Grating (VG), according to which the flow passing through the grating undergoes an angular deviation due to the interaction with inclined slat, is proposed. It models a grating structure with an array of thin inclined slats as a control volume and the flow that hits the slats is forced to deflect by the application of Neumann boundary condition. In other words, inside the region occupied by the grating structure, the flow velocity normal to the slats is set to zero, reproducing the physical effects of the inclined slats. The study also evaluated the wave energy dissipation performance of different configurations of grating structures with inclined slats.

VG was implemented and tested in an in-house projection-based MPS framework (Tsukamoto, et al., 2011), hereinafter referred to as MPS-VG. Nevertheless, it can be easily embedded in other incompressible or weakly compressible particle-based methods, e.g., SPH, incompressible SPH (ISPH) (Cummins & Rudman, 1999; Lo & Shao, 2002), and weakly-compressible MPS (WC-MPS) (Shakibaeinia & Jin, 2010). The main contributions of VG are its simple, efficient, effective, and easily implementable solution. According to the results of MPS-VG simulations, it is highly efficient in processing time and memory usage compared to the conventional fully particle-based MPS modeling. Moreover, it can capture both global and local phenomena involved in multi-scale WSI problems.

This paper is structured as follows: Section 2 highlights the main features of MPS and VG; Section 3 describes the numerical setups for the investigation of VG; Section 4 addresses both convergence and validation of MPS; Section 5 evaluates the computational performance and accuracy of MPS-VG using hydrodynamic loads and green water volumes computed by conventional fully particle-based MPS and SPH simulations as references; Section 6 analyzes the performances of the gratings with inclined slats as a wave damper considering different sizes and positions (vertical or horizontal); finally, Sections 7 and 8, respectively, provide the conclusions and the limitations of the model, as well as suggestions for future developments.

## 2. Numerical method

### 2.1. Moving Particle Semi-implicit method

Navier-Stokes (NS) model, given by the mass conservation and the momentum conservation equations, was considered:

$$\frac{D\rho}{Dt} = -\rho \nabla \cdot \mathbf{u} = 0,  \qquad (1)$$

$$\frac{D\mathbf{u}}{Dt} = -\frac{\nabla P}{\rho} + \nu \nabla^2 \mathbf{u} + \mathbf{f}, \qquad (2)$$

where $\rho$ is fluid density, $\mathbf{u}$ is velocity vector, $P$ is pressure, $\nu$ is kinematic viscosity, $\mathbf{f}$ represents external forces, and $t$ is time.

The differential operators of the governing equations in MPS are replaced by discrete differential operators derived from a weight function. For a given particle $i$, the influence of a neighbor particle $j$ is defined by weight function:

$$\omega_{ij} = \begin{cases} \dfrac{r_e}{\|\mathbf{r}_{ij}\|} - 1 & \|\mathbf{r}_{ij}\| \le r_e \\ 0 & \|\mathbf{r}_{ij}\| > r_e \end{cases}, \qquad (3)$$

where $\|\mathbf{r}_{ij}\| = \|\mathbf{r}_j - \mathbf{r}_i\|$ is the distance between $i$ and $j$ and $r_e$ is the effective radius that limits the range of influence of the neighborhood. The effective radii of $2.1l_0$ and $4.0l_0$ were adopted for the lower-order operators (gradient, divergent, and particle number density) and second-order operators (Laplacian), respectively, where $l_0$ represents the initial distance between particles.

The summation of the weight of all neighbors $\Omega_i$ of particle $i$ is the particle number density ($pnd_i$), which is proportional to the fluid density and computed as

$$pnd_i = \sum_{j \in \Omega_i} \omega_{ij}. \qquad (4)$$

For a scalar function $\phi$ or a vector function $\boldsymbol{\phi}$, gradient, Laplacian, and divergent operators are approximated by:

$$\langle \nabla \phi \rangle_i = \dfrac{d}{pnd^0} \sum_{j \in \Omega_i} \dfrac{\phi_{ij}}{\|\mathbf{r}_{ij}^2\|} \mathbf{r}_{ij} \omega_{ij}, \qquad (5)$$

$$\langle \nabla^2 \phi \rangle_i = \dfrac{2d}{\lambda^0 pnd^0} \sum_{j \in \Omega_i} \phi_{ij} \omega_{ij}, \qquad (6)$$

$$\langle \nabla \cdot \boldsymbol{\phi} \rangle_i = \dfrac{d}{pnd^0} \sum_{j \in \Omega_i} \dfrac{\boldsymbol{\phi}_{ij}}{\|\mathbf{r}_{ij}^2\|} \mathbf{r}_{ij} \omega_{ij}, \qquad (7)$$

where $\phi_{ij} = \phi_j - \phi_i$, $d$ is the number of spatial dimensions, and $pnd^0$ is the initial value of particle number density considering a complete support of the neighboring particles. Constant $\lambda^0$ is a correction parameter so that the variance increase is equal to that of the analytic solution and calculated by:

$$\lambda^0 = \dfrac{\sum_{j \in \Omega_k} \omega_{kj} \|\mathbf{r}_{kj}\|^2}{\sum_{j \in \Omega_k} \omega_{kj}}, \qquad (8)$$

where $k$ is a given particle surrounded by a complete support of isotropic neighboring particle distribution in a Cartesian coordinate system.

### 2.1.1. Boundary conditions

#### Free surface

Neighborhood Particles Centroid Deviation (NPCD), a two-criteria detection technique, was adopted for detecting the free-surface particles (Tsukamoto, et al., 2016). Its first criterion is the original free surface particle detection criterion of MPS:

$$pnd_i < \beta \cdot pnd^0, \qquad (9)$$

where $\beta$ is a scalar parameter empirically determined between 0.80 and 0.99 (Koshizuka & Oka, 1996). If the particle meets the first criterion, the second criterion based on the neighborhood asymmetry is applied to the filtering:

$$\sigma_i = \sqrt{\frac{\sum_{j\in\Omega_i}\|\mathbf{r}_{ij}\omega_{ij}\|}{\sum_{j\in\Omega_i}\omega_{ij}}} > \delta l_0 \quad \text{or} \quad N_i \leq 4, \tag{10}$$

where $\delta \geq 0.2$ (Tsukamoto, et al., 2016) and $N_i$ denotes number of neighbors of particle $i$. If the particle meets both criteria, it is classified as a free surface and its pressure is set to atmospheric pressure. This study adopted the values of 0.98 and 0.2 for parameters $\beta$ and $\delta$, respectively.

**Fixed rigid wall**

The fixed rigid wall was modeled with the use of three layers of particles. The particles of the first layer in contact with the fluid are called wall particles. The pressure of the wall particles is calculated together with the pressure of fluid particles by Pressure Poisson Equation (PPE) (Eq. (18)). The other two layers of particles are denominated dummy particles and used for ensuring the correct calculation of the number density of particles near the wall. Pressure is not calculated for dummy particles. The nonhomogeneous Neumann boundary condition of pressure is applied at rigid walls such that the pressure of a dummy particle j is approximated by:

$$P_j = P_i + \|\mathbf{r}_{ij}\|\frac{\partial P}{\partial n}\bigg|_{\partial\Omega_{wall}}, \tag{11}$$

where Eq. (11) is included in the PPE, see Eq. (18), with the second term $\|\mathbf{r}_{ij}\|\frac{\partial P}{\partial n}\big|_{\partial\Omega_{wall}}$ moved to the right-hand side of the linear system, i.e., the source term.

Regarding forces on fixed rigid walls ($\Omega_W$), the resultant force ($\mathbf{F}_{\Omega_W-MPS}$) in the current MPS is calculated by integrating the pressures ($P_i$) of wall particles $i \in \Omega_W$:

$$\mathbf{F}_{\Omega_W-MPS} = -\sum_{i\in\Omega_W} P_i l_0^2 \mathbf{n}_i, \tag{12}$$

where $\mathbf{n}_i$ is the normal vector at wall particle $i$.

### 2.1.2. Algorithm

The semi-implicit algorithm of MPS is similar to the projection method proposed by Chorin (1967), in which the algorithm is divided into two main parts. In the first, both particle's velocity and position are explicitly predicted considering viscosity and external forces terms of the conservation law of momentum (Eq. (2):

$$\mathbf{u}_i^* = \mathbf{u}_i^t + [\nu\langle\nabla^2\mathbf{u}\rangle_i + \mathbf{f}_i]^t \Delta t, \tag{13}$$

$$\mathbf{r}_i^* = \mathbf{r}_i' + \mathbf{u}_i^* \Delta t . \tag{14}$$

Then, a collision model is applied:

$$\Delta \mathbf{u}_i^* = \begin{cases} \sum_{j \in \Omega_i} \dfrac{(1+\alpha_2)}{\alpha_3} \dfrac{\mathbf{r}_{ij}^* \cdot \mathbf{u}_{ij}^*}{\|\mathbf{r}_{ij}^*\|} \dfrac{\mathbf{r}_{ij}^*}{\|\mathbf{r}_{ij}^*\|} & \|\mathbf{r}_{ij}^*\| \leq \alpha_1 l_0 \text{ and } \mathbf{r}_{ij}^* \cdot \mathbf{u}_{ij}^* < 0 \\ 0 & \text{otherwise} \end{cases}, \tag{15}$$

where $\mathbf{u}_{ij} = \mathbf{u}_j - \mathbf{u}_i$ and values of $\alpha_1 \in [0.8, 1.0]$ and $\alpha_2 \in [0.0, 0.2]$ enhance the spatial stability (Lee, et al., 2011). $\alpha_1 = 0.8$ and $\alpha_2 = 0.2$ were adopted for all cases analyzed. If the neighbor particle is an inner fluid or free-surface particle, then $\alpha_3 = 2$; otherwise, $\alpha_3 = 1$. The contribution of $\Delta \mathbf{u}_i^*$ is added to the particle velocities and positions:

$$\mathbf{u}_i^{**} = \mathbf{u}_i^* + \Delta \mathbf{u}_i^* , \tag{16}$$

$$\mathbf{r}_i^{**} = \mathbf{r}_i^* + \Delta \mathbf{u}_i^* \Delta t . \tag{17}$$

In the second part of the algorithm, the pressure of all particles is implicitly calculated by solving a linear system of PPE. The present study adopted source term Time-scale Correction of Particle-level Impulses (TCPI) proposed by Cheng et al. (2021):

$$\langle \nabla^2 P \rangle_i^{t+\Delta t} - \dfrac{\rho}{\Delta t^2} \alpha_c P_i^{t+\Delta t} = C_s^2 \dfrac{\rho}{l_0^2} \dfrac{pnd^0 - pnd_i^t}{pnd^0} + C_s \dfrac{\rho}{l_0} \langle \nabla \cdot \mathbf{u}^{**} \rangle_i , \tag{18}$$

where $pnd_i^t$ is the number density of particle $i$ at the beginning of instant $t$, $\alpha_c$ is the artificial compressibility coefficient, and $C_s \sim 10\sqrt{gl_0}$ is the speed of propagation of perturbations. Artificial compressibility coefficient $\alpha_c$ was adopted as $10^{-8}$ ms$^2$/kg and the propagation speed of perturbations $C_s$ was calibrated prior to the simulations and is provided in Appendix A.

Finally, velocity and position were updated by Euler integration:

$$\mathbf{u}_i^{t+\Delta t} = \mathbf{u}_i^{**} - \dfrac{\Delta t}{\rho} \langle \nabla P \rangle_i^{t+\Delta t} , \tag{19}$$

$$\mathbf{r}_i^{t+\Delta t} = \mathbf{r}_i^{**} + \left( \mathbf{u}_i^{t+\Delta t} - \mathbf{u}_i^{**} \right) \Delta t . \tag{20}$$

Towards reducing the effect of nonuniform particle distribution and preventing particle clustering, which can lead to unstable computation when attractive forces act between particles ($P_j - P_i < 0$), the gradient model of Wang et al. (2017), rather than the pressure gradient of Eq. (5, was adopted:

$$\langle \nabla P \rangle_i = \left[ \sum_{j \in \Omega_i} \dfrac{\mathbf{r}_{ij}}{|\mathbf{r}_{ij}|} \otimes \dfrac{\mathbf{r}_{ij}^T}{|\mathbf{r}_{ij}|} W_{ij} \right]^{-1} \sum_{j \in \Omega_i} \dfrac{P_j - \hat{P}_i}{\|\mathbf{r}_{ij}\|^2} \mathbf{r}_{ij} W_{ij} , \tag{21}$$

where $\hat{P}_i = \min_{j \in \Omega_i}(P_j, P_i)$ provides exclusively repulsive forces. Notwithstanding, a relevant point is the linear momentum is not conserved in Eq. (21), since the resulting interparticle pressure forces are not anti-symmetric (Khayyer & Gotoh, 2013). Fixed time step $\Delta t$ is set up, satisfying the CFL condition (Courant, et al., 1967):

$$\Delta t \leq \frac{C_r \, l_0}{|\mathbf{u}|_{max}}, \tag{22}$$

where $C_r \in \,]0, 1.0]$ denotes Courant number and $|\mathbf{u}|_{max}$ is maximum flow velocity. For the sake of simplicity, the estimation of the value of $|\mathbf{u}|_{max}$ was based on the upstream initial water column height of the wet dam break experiments (see Section 3.1). $C_r = 0.2$ was used.

## 2.2. Virtual Grating (VG) model

In multi-scale problems, the explicit modeling of small-sized structures or tiny local features by solely wall particles, as in Kashani, et al. (2018), is computationally inefficient and may not be feasible for real-world applications. This is the case of the free-surface flow through gratings with an array of thin inclined slats, which is the focus of the present study. To overcome this drawback and reduce the computational cost of the flow simulations, this study proposes a Virtual Grating (VG) model that replaces the particle-based modeling of the thin slats.

In VG, grating is modeled as a control volume in which the flow velocity is deflected according to the inclination of the grating slats. The flow velocity control is based on Neumann boundary condition, i.e., the flow velocity component normal to the slats is set to zero, reproducing the physical effects of inclined slats. Figure 1 shows the standard wall particle modeling of an inclined slat and the concept of VG applied to a fluid particle before, during, and after the slat had been hit. Angle $0 < |\alpha| < 90°$ is the slat inclination.

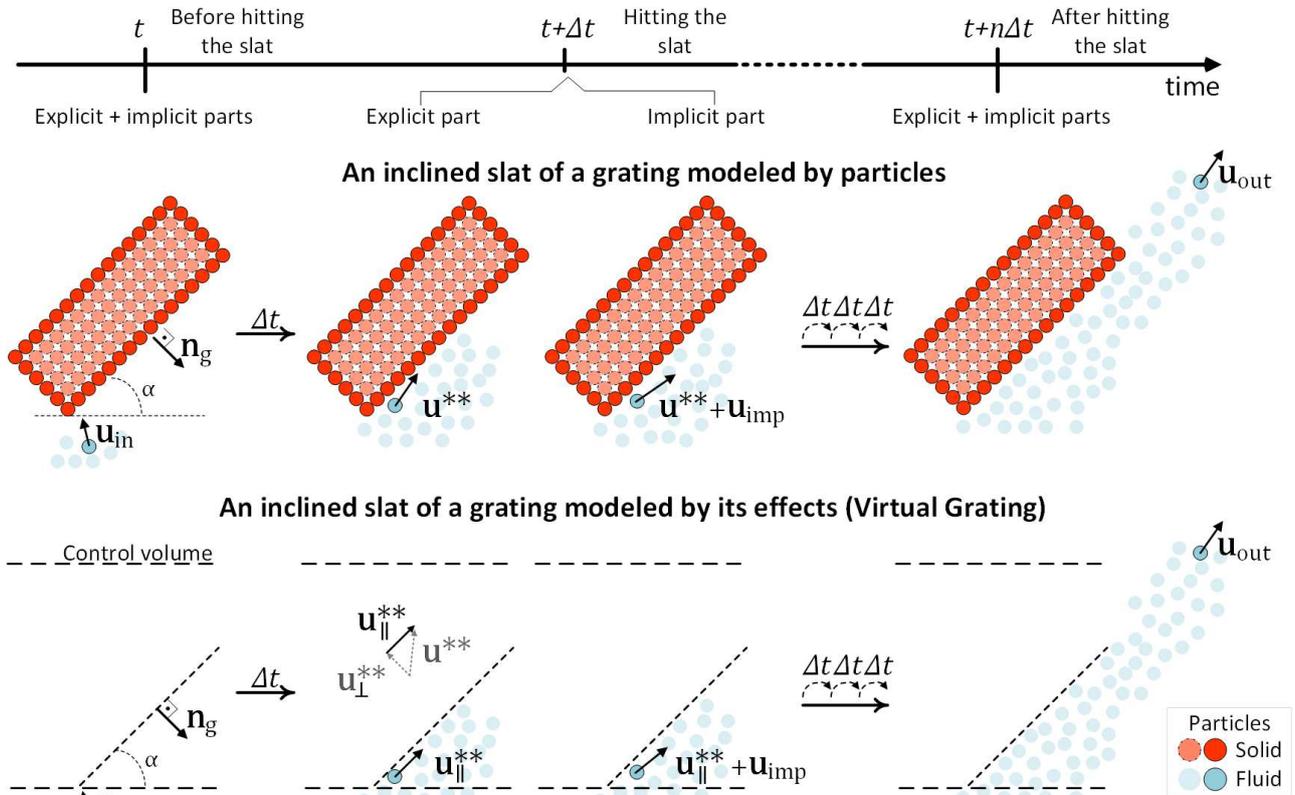

**Figure 1.** Velocity updating of a fluid particle hitting an inclined slat over time. (top) Particle-based grating model with an inclined slat represented by a set of particles and (bottom) Virtual Grating (VG) with the inclined slat modeled by its physical effects.

When the flow enters control volume $\Omega_{VG}$ of VG, e.g., during instant $t + \Delta t$ in Figure 1, its velocity is computed at the explicit time integration part, hereinafter referred to as explicit velocity (see Eq. (16)). The explicit velocity can be decomposed into components normal and tangential to the inclined slat as follows:

$$\mathbf{u}_\perp^{**} = (\mathbf{u}^{**} \cdot \mathbf{n}_g)\mathbf{n}_g, \tag{23}$$

$$\mathbf{u}_\parallel^{**} = \mathbf{u}^{**} - \mathbf{u}_\perp^{**}, \tag{24}$$

where $\mathbf{u}^{**}$ represents the explicit velocity of the fluid particle, $\mathbf{n}_g$ is the normal vector of the inclined slat, and $\mathbf{u}_\perp^{**}$ and $\mathbf{u}_\parallel^{**}$ are, respectively, normal and tangential components of the explicit velocity.

Since the flow velocity normal to the slat must be zero, $\mathbf{u}_\perp^{**} = 0$, only tangential component $\mathbf{u}_\parallel^{**}$ must be considered in the explicit estimation of the fluid particle position (see Eq. (17)). The implicit pressure of the fluid particles inside the control volume is calculated with the use of the explicit estimation of position and tangential velocity $\mathbf{u}_\parallel^{**}$. After the implicit calculation, the velocity of fluid particles inside control volume $\Omega_{VG}$ $\left(\mathbf{u}^{t+\Delta t}\right)_{\Omega_{VG}}$ is then calculated by adding the correction due to pressure gradient term $\mathbf{u}_{\text{imp}}$ to $\mathbf{u}_\parallel^{**}$:

$$\mathbf{u}_{\text{imp}} = -\frac{\Delta t}{\rho} \langle \nabla P \rangle_i^{t+\Delta t}, \tag{25}$$

$$\left(\mathbf{u}^{t+\Delta t}\right)_{\Omega_{VG}} = \mathbf{u}_\parallel^{**} + \mathbf{u}_{\text{imp}}. \tag{26}$$

After a sequence of time steps, the computation of $\left(\mathbf{u}^{t+\Delta t}\right)_{\Omega_{VG}}$ of the fluid particles inside the control volume leads to outflow velocity from the grating, $\mathbf{u}_{\text{out}}$, as shown in Figure 1.

The force estimation in VG is based on Newton's second law of motion as follows:

$$\mathbf{F}_{VG \to \Omega_{VG}} = \sum_{i \in \Omega_{VG}} \left(\frac{d(m\mathbf{u})}{dt}\right)_i, \tag{27}$$

where $\mathbf{F}_{VG \to \Omega_{VG}}$ represents the force the inclined slats, represented by VG, exert on fluid particles $i$ belonging to control volume $\Omega_{VG}$ and $m$ is the mass of the fluid particle, approximated by $m = \rho l_0^2$ in a two-dimensional (2D) domain and $m = \rho l_0^3$ in a three-dimensional (3D) one. Since the particle mass is constant during the simulation and the force is computed after the explicit time integration, Eq. (27) can be simplified by:

$$\mathbf{F}_{VG \to \Omega_{VG}} = m \sum_{i \in \Omega_{VG}} \left(\frac{d\mathbf{u}}{dt}\right)_i \approx m \sum_{i \in \Omega_{VG}} \left(\frac{\Delta \mathbf{u}^{**}}{\Delta t}\right)_i, \tag{28}$$

where $\Delta t$ is the numerical time step and $\Delta \mathbf{u}^{**}$ represents the change caused by VG on the explicit velocity of fluid particle $i$. Since $\Delta \mathbf{u}^{**}$ is equivalent to setting the normal component of the explicit velocity to zero, then:

$$\mathbf{F}_{VG \to \Omega_{VG}} \approx m \sum_{i \in \Omega_{VG}} \left(-\frac{\mathbf{u}_\perp^{**}}{\Delta t}\right)_i. \tag{29}$$

Therefore, the estimation of the force of fluid particles on virtual grating $\mathbf{F}_{\Omega_{VG} \to VG}$ reads:

$$\mathbf{F}_{\Omega_{VG} \to VG} = -\mathbf{F}_{VG \to \Omega_{VG}} \approx m \sum_{i \in \Omega_{VG}} \left( \frac{\mathbf{u}_\perp^{**}}{\Delta t} \right)_i. \quad (30)$$

Assuming the slats are thin and small, the effects of their thickness and the frictional force are neglected in VG.

The following sections address the incorporation of VG in an in-house MPS code (Tsukamoto, et al., 2011) for replacing the conventional particle-based wall modeling of inclined grating slats. The combination of VG model and MPS method is called here MPS-VG.

## 3. Study cases and dimensionless values

### 3.1. Study cases

The wet dam break experiment conducted by Hernandez-Fontes, et al. (2020) was adopted for evaluations of the performances of MPS and VG. From a practical engineering point of view, the wet dam break model reasonably predicts the nonlinearities of geophysical and free-surface hydraulics phenomena such as urban flooding, tsunami-like bore, and isolated green water flow on deck (Aureli, et al., 2023). It consists of a gate that separates upstream and downstream water columns of heights $h_0$ and $h_1$, respectively, and a fixed structure of length $L_{FS} = 0.195$m and height $H_{FS} = 0.15$m in the downstream end, as illustrated in Figure 2. Table 1 shows the physical parameters of the simulations. The duration of the simulated wet dam break events was 2.0s.

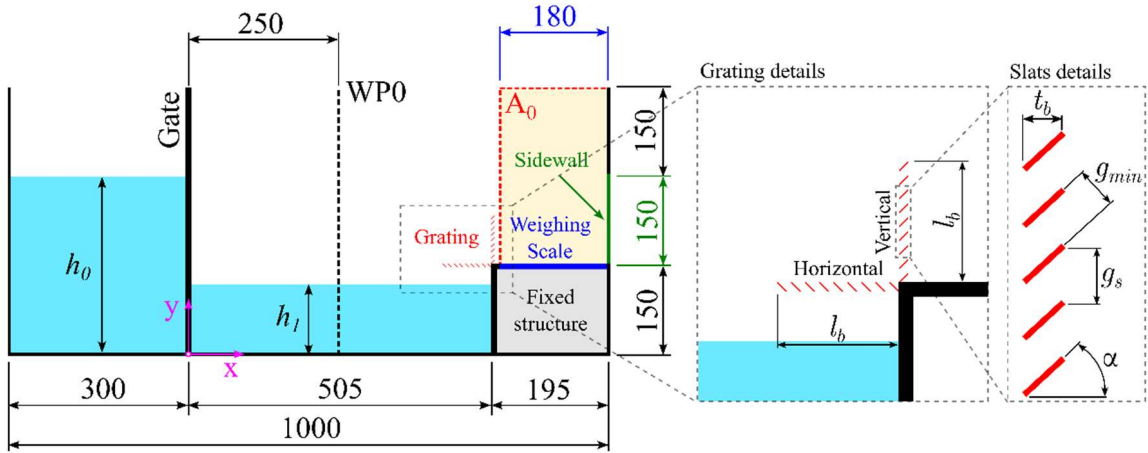

**Figure 2.** Initial wet dam breaking model setup and gratings with thin inclined slats (dimensions in mm). $A_0$ stands for the region adopted for computing the water volume above the fixed structure.

**Table 1.** Physical parameters of the simulations.

| | |
|---|---|
| Fluid density $\rho$ [kg/m³] | 1000 |
| Fluid kinematic viscosity $\nu$ [m²/s] | $10^{-6}$ |
| Gravitational acceleration $g$ [m/s²] | 9.81 |

2D Simulations were conducted considering water column heights $h_0$ and $h_1$, summarized in Table 2, following the main dimensions and four test conditions available in Hernández-Fontes, et al. (2020).

Table 2. Initial water column heights for the four simulated conditions.

| Condition | $h_0$ [m] | $h_1$ [m] |
|---|---|---|
| C1 | 0.17 | 0.12 |
| C2 | 0.20 | 0.12 |
| C3 | 0.24 | 0.12 |
| C4 | 0.30 | 0.12 |

Due to a lack of experimental tests covering the interaction between free-surface flow and grating with inclined slats, conventional fully particle-based simulations using MPS and Weakly Compressible SPH (WCSPH) were adopted as references for evaluating the performance of MPS-VG. In MPS and SPH simulations, the entire computational domain was modeled with the use of particles and the thin inclined slats are represented as an array of tiny fixed rigid bodies defined by particles. As reported in Amaro Jr. et al. (2021) and Hashimoto et al. (2022), both MPS and SPH can provide accurate results of hydrodynamic forces.

The same in-house MPS code (Tsukamoto, et al., 2011) used for the MPS-VG implementation was adopted for the MPS simulations, whereas the Open-source code DualSPHysics v5.0.5 (Domínguez, et al., 2022) was used for the SPH simulations. DualSPHysics is based on a weakly compressible SPH model for the fluid phase. The input files used in DualSPHysics v5.0.5 are provided in the supplementary data so that readers can reproduce the WCSPH simulations.

According to previous studies on breakwaters (Buchner, 2002; Lee, et al., 2012) and riser balconies (Rosetti, et al., 2019), the main dimensions of the prototype are approximately 4.0m. Since the scale factor of the experiments in Hernández-Fontes, et al. (2020) is around 1:60, gratings of lengths $l_b = 0.040 m$ (2.4m on prototype scale) and $0.075 m$ (4.5m on prototype scale) placed horizontally (horizontal grating) and vertically (vertical grating) are considered herein. The gratings have $t_b = 4.243$mm thickness (0.255m on prototype scale) and the inclination of the slats is $\alpha = 45°$ with $g_s = 8.657$mm spacings (0.519m on prototype scale) between them (see slats details in Figure 2).

The nomenclature adopted follows the [*grating position*]-[*grating length*]-[*initial water column height*] pattern. Table 3 provides the abbreviations of the grating and initial water column characteristics. For instance, VT-L75-C3 denotes the case with a vertical grating of length $l_b = 0.075$m in condition C3.

Table 3. Nomenclature of simulated cases.

| | | |
|---|---|---|
| Grating position | Horizontal | HZ |
| | Vertical | VT |
| Grating length | $l_b = 0.040$ m | L40 |
| | $l_b = 0.075$ m | L75 |
| Initial water column height | $h_0 = 0.17$ m | C1 |
| | $h_0 = 0.20$ m | C2 |
| | $h_0 = 0.24$ m | C3 |
| | $h_0 = 0.30$ m | C4 |

### 3.2. Dimensionless values

The following dimensionless values for water elevation $\eta^*$, time $t^*$, force $F^*$, and respective impulse $I^*$ were adopted for all results:

$$\eta^* = \frac{\eta}{h_0}, \tag{31}$$

$$t^* = t\sqrt{\frac{g}{h_0}}, \tag{32}$$

$$F^* = \frac{F_{2D}W}{\rho g h_0 L_{FS}^2}, \tag{33}$$

$$I^* = \frac{I_{2D}W}{\rho g^{0.5} h_0 L_{FS}^{2.5}}, \tag{34}$$

where $L_{FS} = 0.195$m is the length of the fixed structure (see Figure 2), $W = 0.335$m is the tank width in the experiments, and $F_{2D}$ is the hydrodynamic force computed by 2D simulations. Impulse $I_{2D}$ is a meaningful representation of the hydrodynamic impact loads between instants $t_0$ and $t_f$ and can be expressed as

$$I_{2D} = \int_{t_0}^{t_f} F_{2D} dt. \tag{35}$$

For the sake of simplicity, the hydrodynamic loads on the weighing scale were adopted as estimated loads on the fixed structure. Similarly, the hydrodynamic loads on the 150 mm height downstream vertical sidewall were used for estimating the loads on the installations to be protected. The weighing scale and the vertical sidewall are displayed in Figure 2 in blue and green colors, respectively.

In the MPS-VG simulations, the hydrodynamic forces on the inclined grating slats were obtained by Eq. (30) and the forces on the weighing scale and sidewall were computed by Eq. (12), whereas Eq. (12) is also used to calculate all forces obtained in the conventional MPS simulations. Using DualSPHysics v5.0.5 (Domínguez, et al., 2022) in the SPH simulations, the forces were determined by the summation of the momentum variation of neighboring fluid particle $f \in \Lambda_i$ inside the compact support of boundary wall particle $i$:

$$\mathbf{F}_{\Omega_W-\text{SPH}} = \sum_{f \in \Lambda_i} m_f \mathbf{a}_{if}, \tag{36}$$

where $\mathbf{F}_{\Omega_W-\text{SPH}}$ is the force exerted by the fluid against a rigid wall, $m_f$ is the mass of fluid particle $f$, and $\mathbf{a}_{if}$ is the difference in acceleration between $i$ and $f$.

Besides hydrodynamic forces, the water-on-deck flow is useful for the estimation of loads (Silva & Rossi, 2014) and risk to personnel or equipment on the deck and its critical parameters are height and velocity. However, the flow downstream of the grating with inclined slats is usually a mixture of jet flow, breaking waves, and splash and, under such conditions, the measurement of water height is not feasible. As an alternative, 2D water volume (in square meter m²) $V_{2D}$ was analyzed within the highlighted region $A_0 = 0.054$m² above the fixed structure, as illustrated in Figure 2. The water volume is related to the mean height of the water on deck and dimensionless average values of water volume $V^*$ were computed as

$$V^* = \frac{\int_{t_0}^{t_f} V_{2D} dt\, W}{h_0 L_{FS}^2 (t_f - t_0)}. \tag{37}$$

## 4. Convergence and validation of MPS

### 4.1. MPS spatial convergence

The numerical convergence of MPS was evaluated using the most severe condition, C4, described in Section 3 (see Table 2). Gratings were not considered in this numerical convergence study. Resolution, initial distance between particles, number of particles, time step, processing time, and speed of propagation of perturbations $C_s$ adopted for the simulations are provided in Table 4.

**Table 4.** Numerical parameters, number of particles, and processing time adopted in the convergence study.

| Resolution $h_1/l_0$ | 16 | 32 | 64 | 128 | 256 |
|---|---|---|---|---|---|
| Initial distance between particles $l_0$ [mm] | 7.5 | 3.75 | 1.875 | 0.9375 | 0.46875 |
| Number of particles | 3450 | 12260 | 45870 | 177460 | 697490 |
| Speed sound $C_S$ [m/s] | 2.0 | 1.4 | 1.0 | 0.7 | 0.5 |
| Time step $\Delta t$ [ms] | 0.50 | 0.20 | 0.10 | 0.05 | 0.02 |
| Processing time* [h] | 0.01 | 0.05 | 0.45 | 3.8 | 45.92 |

*Intel® Xeon® Processor E5-2680 v2 2.80GHz, 10 Cores (20 Threads).

Figure 3 displays comparisons between the numerical results from MPS and five experimental repetitions (Hernández-Fontes, et al., 2020). As reported in Hernández-Fontes et al., (2020), measurements started when a trigger was activated to release the gate and, since it occurred between instants $0.4s < t < 0.6s$ after the trigger activation, $t = 0.5s$ was taken as a reference for the beginning of the dam break event.

Figure 3 (a) shows the time series of the vertical force on the weighing scale. The numerical results are raw data obtained from the MPS simulations performed in this study, whereas filtered experimental data were provided by Hernández-Fontes, et al. (2020). Although the first peaks of computed force occurred slightly earlier, the overall trend of numerical and experimental results shows reasonably good agreement. Since the numerical modeling of the interaction between fluid and sluice gate considering opening velocity, roughness of the wall, and fluid viscosity is challenging, an instantaneous dam-break was taken into account in the present study. Such a simplification in the numerical model may be the reason for the time lag between numerical and experimental results. Discussions on the effects of gate opening in the context of particle-based methods can be found in (Jandaghian & Shakibaeinia, 2020; Areu-Rangel, et al., 2021). Another reason may be attributed to the simplification of 2D domains, which cannot take into account the 3D effects observed in the experimental results. The MPS results over-predicted the impact forces, as in the SPH results shown in Fig. 5(h) in Areu-Rangel et al. (2021). Such differences can also be partially attributed to an "air cushion" effect, which was not modeled in the numerical simulations because the air phase was neglected.

Compared to the computed forces, the impulse is less prone to oscillations and enables the verification of trends among different model resolutions. Dimensionless impulse $I^*$ (Eq. (34)) of the vertical force on the weighing scale was used in the convergence analysis. Following the Grid Convergence Index (GCI) method presented in Roache (1998; 2009), the numerical uncertainty was estimated in the computed impulse values. The difference between the values obtained by 3 grid resolutions $\epsilon_{21}$ and $\epsilon_{32}$ were used - subscripts 1, 2, and 3 refer to fine, mid, and coarse resolution grids, respectively. The asymptotic ratio was computed as $Ar = r^p \text{GCI}_{12}/\text{GCI}_{23}$, where $r$ is the grid refinement ratio equal to 2.0, $p$ is the order of convergence $p = \ln\left(\frac{I_3^* - I_2^*}{I_2^* - I_1^*}\right)/\ln(r)$, GCI is the grid refinement index computed as $\text{GCI}_{12} = \frac{F_S|\varepsilon_{12}|}{(r^p - 1)}$, $F_S$ is a safety factor equal to 1.25, and $\varepsilon$ is relative error $\varepsilon_{12} = \frac{I_2^* - I_1^*}{I_1^*}$.

Figure 3 (b) shows the impulse, computed over the 0 to 1.5s time interval, as a function of the model resolution, and a table with the values of convergence. Impulse converges to $I^* \approx 2.5$ for resolutions $h_1/l_0 \geq 128$. In this case, the results show a monotonic convergence, since $\epsilon_{21}/\epsilon_{32} = 0.33$ is between 0 and 1 and asymptotic ratio $Ar = 1.01$ very close to 1 indicates the solutions are well within the asymptotic range of convergence. Therefore, the initial distance between particles $l_0 = 0.9375$mm ($h_1/l_0 = 128$) was chosen for the MPS validation study.

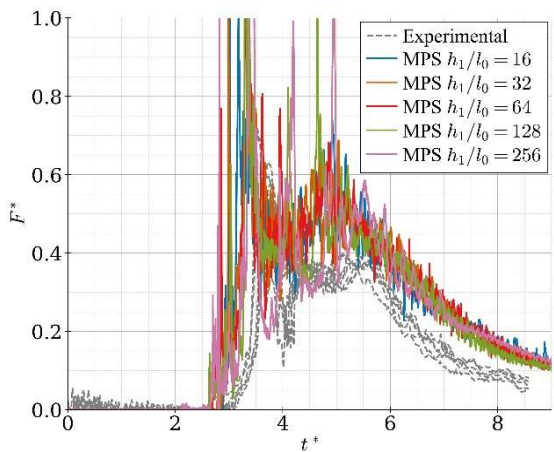
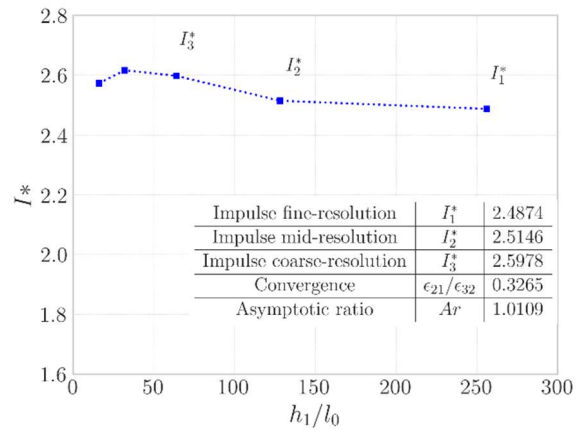

(a) Vertical force on the weighing scale  (b) Impulse of force on the weighing scale

**Figure 3.** Time series of (a) vertical force on weighing scale for five experimental repetitions (Hernández-Fontes, et al., 2020) and MPS using five resolutions for condition C4 ($h_0 = 0.3$m). (b) Impulses of forces computed by MPS.

### 4.2. MPS Validation

The MPS validation was evaluated by comparing numerical and experimental results of water elevations at the wave probe WP0 and vertical forces on the weighing scale. Conditions C1 to C4 (Table 2) with no gratings were considered. The computed forces and respective impulses were multiplied by tank width $W = 0.335$ m for comparing the 2D numerical results with the 3D experimental ones.

Snapshots of the experiments and simulations at three representative instants of the green water event for conditions C1 and C4 are provided in Figure 4 and Figure 5, respectively. The snapshots show the smooth pressure fields obtained by MPS, which predicts well the main hydrodynamic behaviors of both calm and severe conditions.

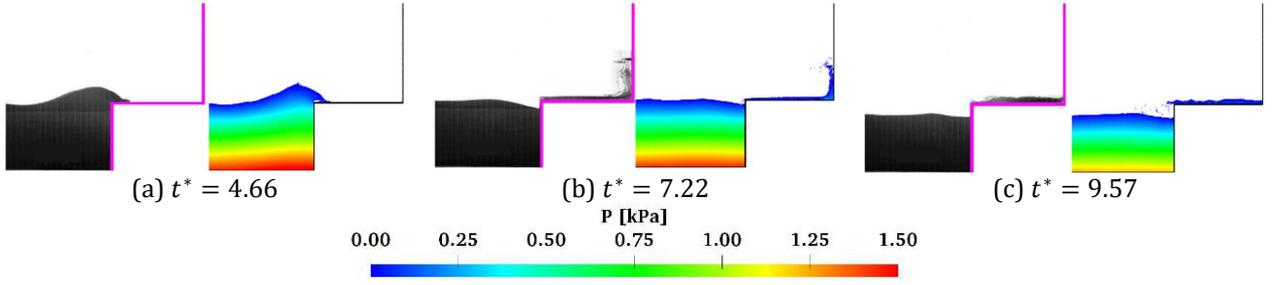

**Figure 4.** Snapshots of the wet dam breaking event with no grating at three instants in the experiment (Hernández-Fontes, et al., 2020) and simulation for C1 ($h_0 = 0.17$m). The color scale represents pressure magnitude.

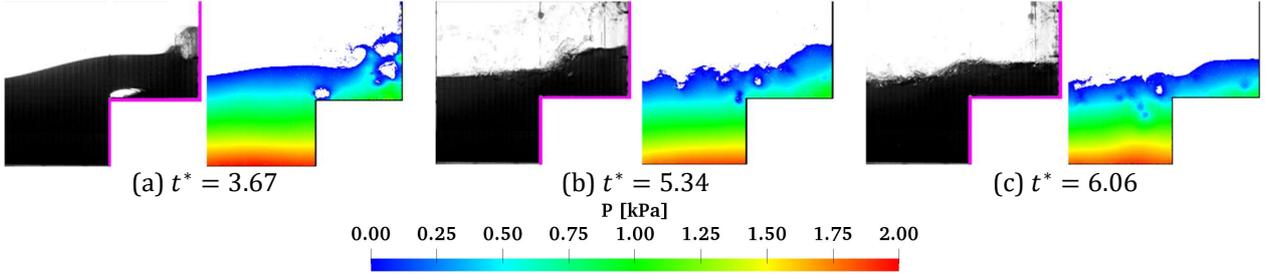

**Figure 5.** Snapshots of the wet dam breaking event with no grating at three instants in the experiment (Hernández-Fontes, et al., 2020) and simulation for C4 ($h_0 = 0.3$m). The color scale represents pressure magnitude.

The MPS accuracy was analyzed with the use of three statistical metrics, namely, normalized centered root-mean-square difference ($E_n$), see Eq. (42), standard deviation ($\sigma_n$), and correlation ($R$), defined as

$$\sigma_{n,A} = \frac{\sqrt{\frac{1}{N}\sum_{i=n}^{N}(A_i - \overline{A})^2}}{\sigma_B}, \qquad (38)$$

$$R_A = \frac{\sum_{i=n}^{N}[(A_i - \overline{A}_i)(B_i - \overline{B})]}{N\sigma_A \sigma_B}, \qquad (39)$$

where $A$ and $B$ refer to numerical and experimental data (here, experimental repetition number 3), respectively, and $N$ is the number of samples. Standard deviation $\sigma$ is given by Eq. (43) and mean values are represented by an overhead bar. As shown in Figure 6(a), the water elevation at WP0 computed for C1 and C2 conditions match very well with the experimental ones. Moreover, a reasonably good agreement was obtained for C3 and C4, in which the maximum water elevation achieved by MPS occurred earlier. The time lag can be explained by the effects of the gate opening, which was not modeled in the simulations. As expected, it increases as the upstream water column increases.

The Taylor diagram (Taylor, 2001) in Figure 6(b) illustrates the three statistical metrics. The ranges of values obtained are $0.85 < R < 0.98$, $0.22 < E_n < 0.57$, and $\sigma_n$ close to 1.0. The values, with $R$ and $\sigma_n$ close to 1 and $E_n$ near to 0, indicate a good MPS prediction accuracy for the wave elevation.

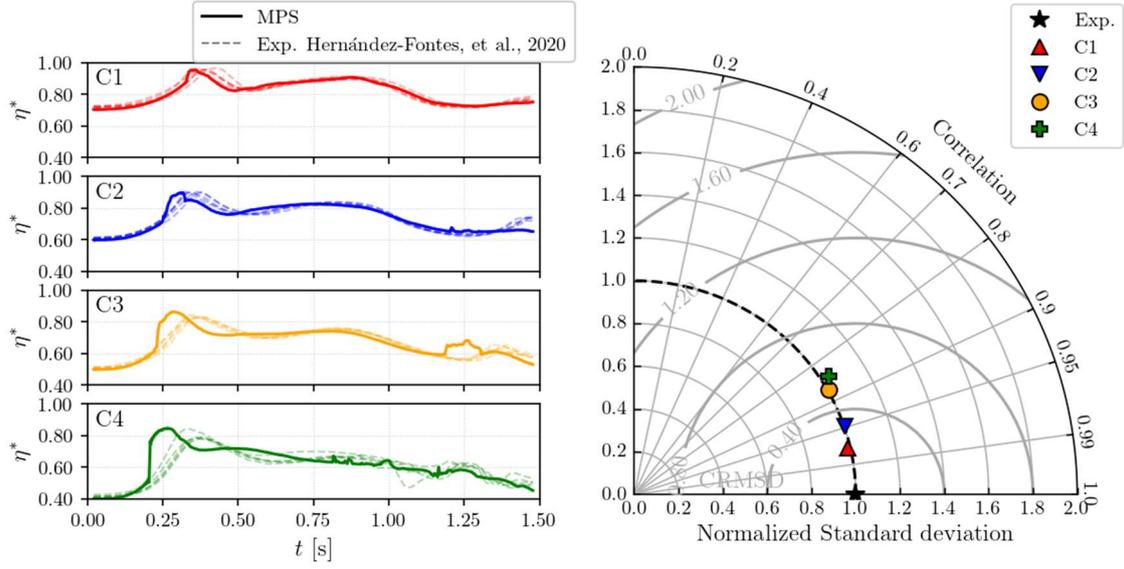

(a) Water elevation at WP0          (b) Taylor diagram of water elevation

**Figure 6.** (a) Water elevation ($\eta^*$) at WP0 for five experimental repetitions (Hernández-Fontes, et al., 2020) and numerical results of MPS with resolution $h_1/l_0 = 128$ ($l_0 = 0.9375$mm) for C1, C2, C3, and C4. (b) Taylor diagram of three statistical metrics of water elevations (the 3$^{rd}$ experimental repetition is used as a reference).

According to Figure 7(a), the computed and experimentally measured vertical forces on the weighing scale show similar hydrodynamic impact events. However, after the first peak, the MPS simulations slightly overestimated the forces, presumably due to some numerical approximations or 3D effects that may have been neglected in the 2D model. Such errors were also observed in the SPH results in Fig. 5 in Areu-Rangel et al. (2021).

Figure 7(b) shows the Taylor diagram with the three statistical metrics. The $0.75 < R < 0.91$, $0.71 < E_n < 0.75$, and $1.1 < \sigma_n < 1.5$ ranges of values indicate reasonably good MPS accuracy for force prediction.

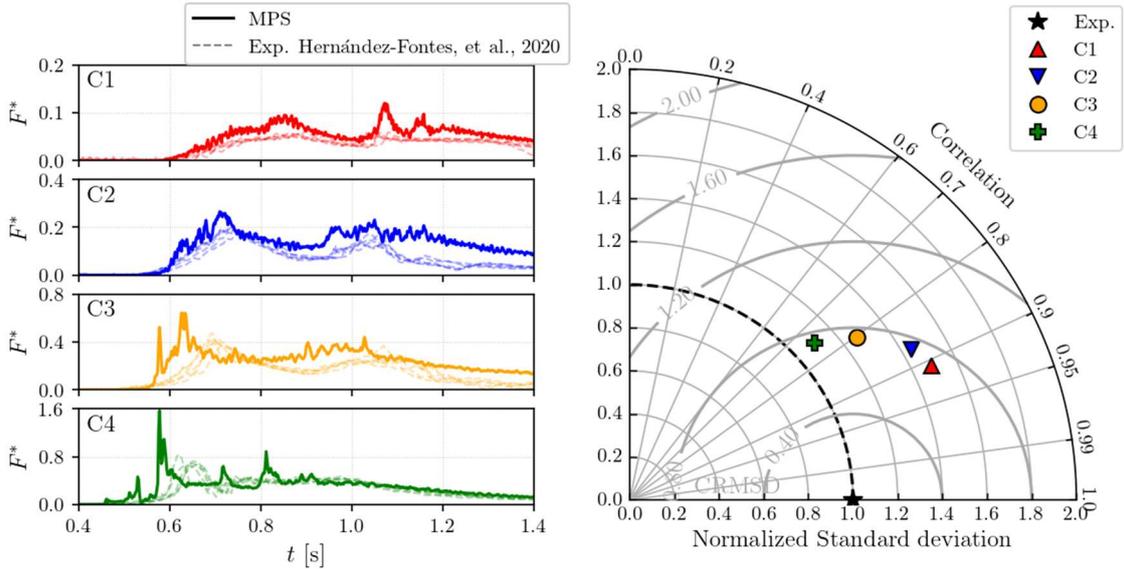

(a) Vertical force on the weighing scale          (b) Taylor diagram of force

**Figure 7.** (a) Vertical force ($F^*$) on the weighing scale for five experimental repetitions (Hernández-Fontes, et al., 2020) and numerical results of MPS with resolution $h_1/l_0 = 128$ ($l_0 = 0.9375$mm) for C1, C2, C3, and C4. Force-axis scales are different for better visualization. (b) Taylor diagram of the three statistical metrics of the vertical forces on the weighing scale (the 3$^{rd}$ experimental repetition was used as a reference).

Two other metrics provided in González-Cao, et al. (2019), namely, Skill Index ($SI$) and Accuracy Index ($AI$) were considered:

$$SI_A = \frac{4(1 + R_A)^4}{\left(\sigma_{n,A} + \frac{1}{\sigma_{n,A}}\right)^2 (1 + R_0)^4}, \qquad (40)$$

$$AI_A = |1 - R_A^2| + E_{n,A}. \qquad (41)$$

where $R_0$ is the maximum correlation attainable for the model (equal to 1).

Ideally, values $SI = 1$ and $AI = 0$ indicate coincident numerical and experimental results. Table 5 summarizes the values of the two metrics for wave elevations ($SI_w$ and $AI_w$) and vertical forces ($SI_f$ and $AI_f$) related to the MPS results and the 3$^{rd}$ experimental repetition. Index values for wave elevation in the $0.72 < SI_w < 0.95$ and $0.27 < AI_w < 0.86$ ranges and for the vertical forces in the $0.58 < SI_f < 0.71$ and $0.89 < AI_f < 1.18$ ranges suggest reasonably well predictions.

**Table 5.** Skill Index (SI) and Accuracy Index (AI) for the computed water elevations ($\eta^*$) and vertical forces ($F^*$) on the weighing scale, considering the 3$^{rd}$ experimental repetition as a reference.

| Condition | Wave elevation | | Force | |
|---|---|---|---|---|
| | Skill Index ($SI_w$) | Accuracy Index ($AI_w$) | Skill Index ($SI_f$) | Accuracy Index ($AI_f$) |
| C1 | 0.95 | 0.27 | 0.71 | 0.89 |
| C2 | 0.90 | 0.42 | 0.68 | 0.98 |
| C3 | 0.77 | 0.74 | 0.63 | 1.11 |
| C4 | 0.72 | 0.86 | 0.58 | 1.18 |

## 5. Numerical accuracy and computational performance of MPS-VG

### 5.1. Spatial convergence of MPS-VG

According to the convergence and validation analysis detailed in Section 4, the initial distance between particles used in the MPS simulations of the wet dam break with no gratings should be smaller than $l_0 = 0.9375$mm. Nevertheless, in the wet dam break with gratings, a finer resolution is required for both conventional fully particle-based MPS and SPH simulations to capture the details of the interaction between slats and fluid. Therefore, the initial distance between particles of $l_0 = 0.234375$mm ($h_0/l_0 \geq 512$) was adopted to ensure the minimum gap ($g_{min} = 6.14$mm) between the grating slats would have at least 20 $l_0$, as depicted in Figure 8. The relation between gap dimension and initial particle distance follows the lowest resolution recommended by Amaro Jr. et al. (2019) for the modeling flow through the openings.

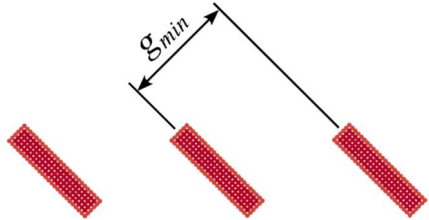

**Figure 8.** Detail of the inclined slats modeled by particles with an initial distance of $l_0 = 0.234375$mm ($h_0/l_0 \geq 512$) and $g_{min} = 6.14$mm minimum gap between the slats, i.e., resolution $g_{min}/l_0 \geq 20$ was required to assure an adequate modeling of the flow through the openings.

The spatial convergence of the MPS-VG modeling was analyzed with the use of the wet dam break case VT-L75-C4 (see Table 3). The computed dimensionless impulses of the force on the inclined slats grating, calculated in the $[0s, 2.0s]$ time interval (see Eq. 33), and five resolutions provided in Table 4 were considered.

Figure 9(a) displays the results filtered by a low-pass filter with an $f_c = 100Hz$ cutoff frequency. Overall, the forces show a similar trend, although the magnitude of the first force peak is higher as the resolution ($h_1/l_0$) increases, which is reasonable, considering the localized and impulsive nature of hydrodynamic impact loads.

According to Figure 9 (b), the impulse converges to $I^* \approx 0.43$ for resolutions $h_1 / l_0 \geq 64$. The results show an oscillatory convergence also associated with the negative value of $\epsilon_{21}/\epsilon_{32} = -0.65$. Moreover, asymptotic ratio $Ar = 0.97$, close to 1, shows good convergence. Since the fine-resolution model ($h_1 / l_0 = 256$) does not improve accuracy substantially, the distance between particles $l_0 = 0.9375mm$ ($h_1 / l_0 \geq 128$) was considered adequate for the simulations of wet dam-break with grating by MPS-VG. The model resolution is the same adopted for the MPS simulations of the cases with no grating (Section 4.1). In other words, the tiny grating slats can be modeled by MPS-VG using the resolution required for the global flow event with no concern over the details of the tiny structural elements.

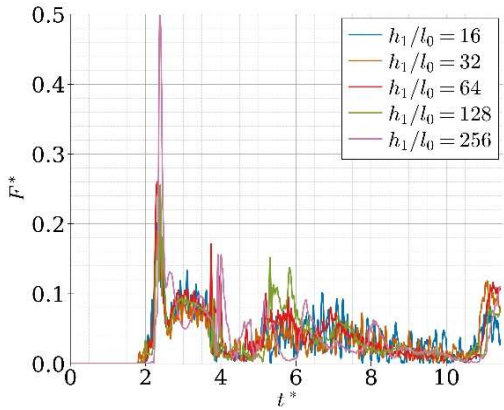
(a) Force normal to the slats obtained by MPS-VG (filtered data)

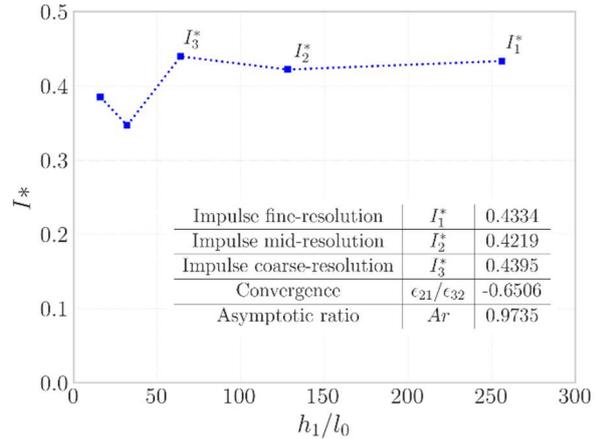
(b) Impulse of force on the weighing scale

**Figure 9.** (a) Time series of the force normal to the slats and (b) respective impulse on the grating obtained by MPS-VG for VT-L75-C4 using five resolutions. Force filtered by a low-pass filter ($f_c = 100Hz$).

### 5.2. Numerical parameters

The numerical parameters adopted for the MPS-VG, MPS, and SPH simulations are listed in Table 6. DualSPHysics uses an adaptive time step restricted by advection and diffusion stability criteria (see Eq. (35) in Domínguez et al. (2022)). Moreover, for condition C4, only the results from MPS-VG and SPH are shown because of numerical instabilities in MPS.

Table 6. Numerical parameters of MPS-VG, MPS, and SPH.

| Parameters | MPS-VG | MPS | SPH |
|---|---|---|---|
| Particles distance $l_0$ [mm] | 0.9375 | 0.234375 | 0.234375 |
| Radius of support $r_e$ | lower order op: $2.1 l_0$<br>second order op: $4.1 l_0$ | lower order op: $2.1 l_0$<br>second order op: $4.1 l_0$ | $2h = 2\sqrt{2} l_0$ |
| Time step [ms] | 0.05 | 0.01 | Adaptative |
| Simulation time [s] | 2.0 | 2.0 | 2.0 |
| Free-surface condition | $\beta = 0.98, \delta = 0.20$ | $\beta = 0.98, \delta = 0.20$ | - |
| Collision coefficients | $\alpha_1 = 0.85, \alpha_2 = 0.2$ | $\alpha_1 = 0.85, \alpha_2 = 0.2$ | - |
| Reference density $\rho_0$ [kg/m³] | - | - | 1000 |
| Polytropic constant $\gamma$ | - | - | 7 |
| Coefficient of sound | - | - | 20 |
| Artificial compressibility coefficient $\alpha_c$ [ms²/kg] | $10^{-8}$ | $10^{-8}$ | - |
| Speed of sound $C_S$ [m/s] | 0.7 | 0.35 | - |

### 5.3. Horizontal grating

Figure 10 and Figure 11 show the evolution of the wave impact against the grating for HZ-L75-C3 at three instants obtained by MPS and MPS-VG, respectively. The color scale on the fluid particles is related to their pressure.

At $t^* = 2.557$, the wave runup along the freeboard reaches the grating and is deflected, as shown in Figure 10(a) and Figure 11(a). Part of the fluid merges with the plunging wave (see Figure 10(b) and Figure 11(b)). Afterward, the incident plunging wave impacts the weighing scale and hits the downstream sidewall, as depicted in Figure 10(c) and Figure 11(c).

Overall, both pressure fields computed by MPS and MPS-VG are similar. The main discrepancy between the flow profiles is the jet flow formed downstream of the grating, as depicted in Figure 10(a) and Figure 11(a). A single jet flow created in the MPS-VG simulation is presumed to occur under ideal conditions involving very thin and small-spaced slats. On the other hand, jet flows were formed in the gaps between the thin inclined slats in the MPS simulation, demanding finer resolutions for a better reproduction. The vena contract of the jet flow passing through the gaps can also be visualized in Figure 10(a). Whereas VG forces a complete flow deflection following slats angle $\alpha = 45°$, the jet flow angle obtained by the MPS simulation is smaller than $\alpha$ due to a relatively low solidity ratio of the grating, i.e., $t_b = \tan \alpha / g_s = 0.49$, so that the flow deflection is not fully developed.

As highlighted in the zoom-in in Figure 11(a), in case of a violent hydrodynamic impact, the flow deflected by VG leads to an abrupt change in particle number density $pnd_i$, causing discontinuities of pressure in the control volume of VG. Other discrepancies of the wave-breaking profiles after $t^* = 4.155$ (Figures (b)-(c) and Figure 11(b)-(c)) are acceptable due to the chaotic nature of the stage, as previously reported by Wei et al. (2018) in a systematic experimental and numerical investigation.

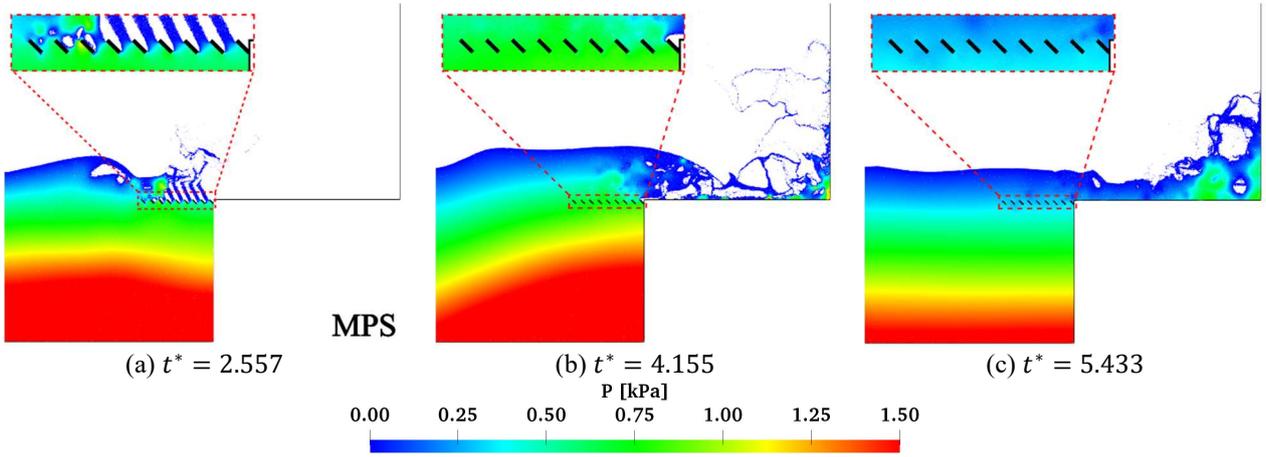

**Figure 10.** Snapshots of wet dam breaking event on a fixed structure for HZ-L75-C3 at three instants obtained by MPS simulations. The color scale represents pressure magnitude.

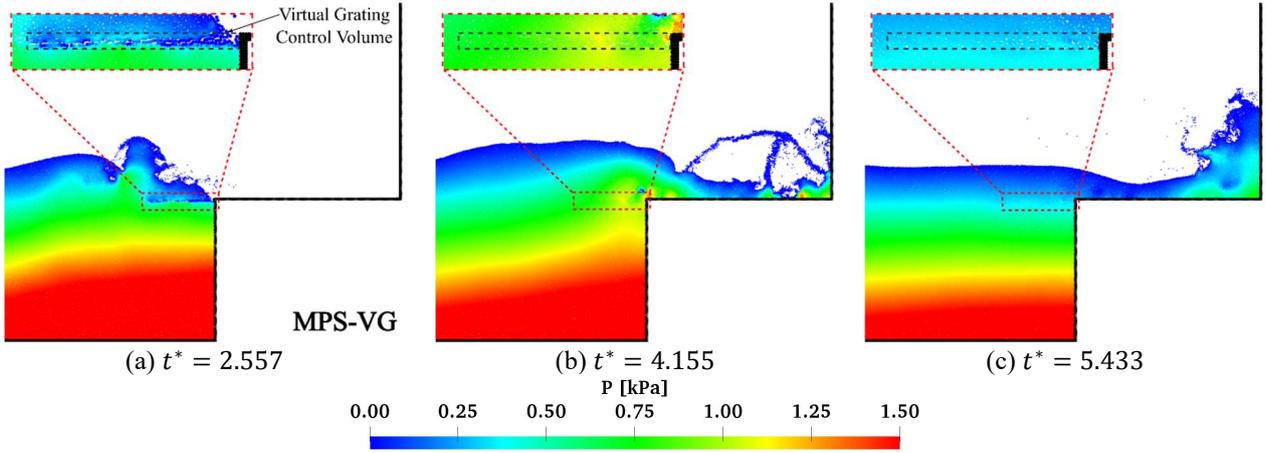

**Figure 11.** Snapshots of wet dam breaking event on a fixed structure for HZ-L75-C3 at three instants obtained by MPS-VG simulations. The color scale represents pressure magnitude.

### 5.3.1. Forces on horizontal gratings, weighing scale, and sidewall

Figure 12 shows the time series of the magnitudes (Euclidean norm) of the normal forces on the slats of horizontal gratings of length $l_b = 0.040$m and 0.075m under conditions C1 and C3.

As shown in Appendix B, MPS-VG computes the force on the grating using Eq. (30), which leads to high-frequency noises, and a low-pass filter with $f_c = 100$Hz cutoff frequency was applied for a better visualization. However, the raw data of the forces were used for the calculation of all related values, e.g., impulses and discrepancies, reported in the following sections. In Figure 12, the time series of the forces computed by MPS and SPH are raw data.

The results for the less severe condition (C1) are provided in Figure 12(a1) and (b1). A sudden increase occurs at $t^*{\sim}3.5$ due to wave runup impact, and a second peak caused by the incident plunging wave follows at $t^*{\sim}4$, which is more evident in the MPS results. Another peak occurs at $t^*{\sim}5.5$ to $t^*{\sim}6.5$ due to plunging wave breaking. The flow returns to the weighing scale and the force increases after $t^*{\sim}10.0$. A second wave runup along the freeboard then hits the grating and generates the force peak after $t^*{\sim}14$.

For the more severe condition (C3), shown in Figure 12(a2) and (b2), the runup along the freeboard hits the horizontal gratings and leads to the force peak at $t^*\sim 2$. As the flow returns, a second load peak on the gratings occurs at $t^*\sim 8$. A third peak occurs at $t^*\sim 12$ due to second wave runup along the freeboard.

Figure 13 and Figure **14**, respectively, display the time series of the forces on the weighing scale and sidewall protected by the horizontal gratings. All results are shown in raw data. According to Figure 13(a1) and (b1) and Figure 14(a1) and (b1), very small forces on the weighing scale and sidewall occur in relatively calm condition C1. For the more severe condition (C3), the first peak of force on the weighing scale occurs at $t^*\sim 4$ due to the incident plunging wave, followed by a second peak at $t^*\sim 6.5$ caused by the returning flow (see Figure 13(a2) and (b2)). The force on the sidewall increases during the flow runup at $t^*\sim 3$, as depicted in Figure 14(a2) and (b2).

The computed time series show the severity of the wave impact condition and the length of the horizontal grating does not affect the overall behaviors and the succession of events of the wave-grating interaction phenomena. Moreover, increasing the horizontal grating from 0.040 m to 0.070 m, the reduction in the hydrodynamic loads on the weighing scale and the sidewall protected by the horizontal grating is relatively small.

In general, the force time series obtained by MPS-VG agree well with those computed by MPS, indicating VG can capture all the relevant events of the highly nonlinear wave impact on gratings with inclined slats. On the other hand, the SPH results also show similar behaviors despite some high-frequency oscillations, which might be due to acoustic perturbations in the pressure field associated with the weakly compressible formulation of SPH.

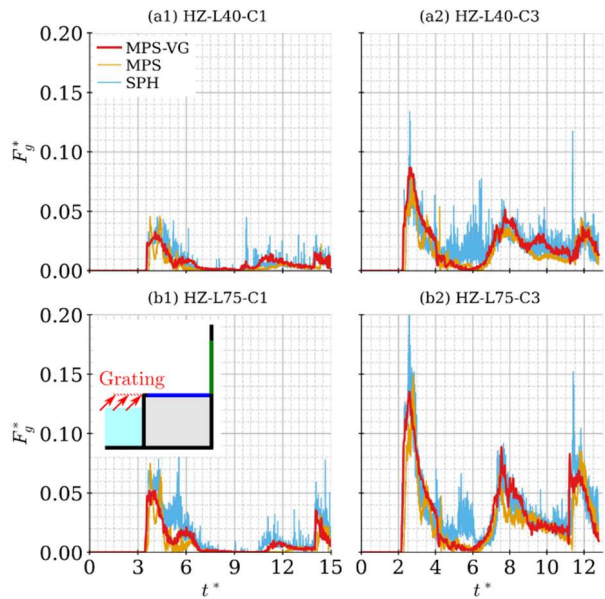

**Figure 12.** Normal forces on the inclined slats of (a1) HZ-L40-C1, (a2) HZ-L40-C3, (b1) HZ-L75-C1, and (b2) HZ-L75-C3. MPS-VG force filtered by a low-pass filter ($f_c = 100 Hz$).

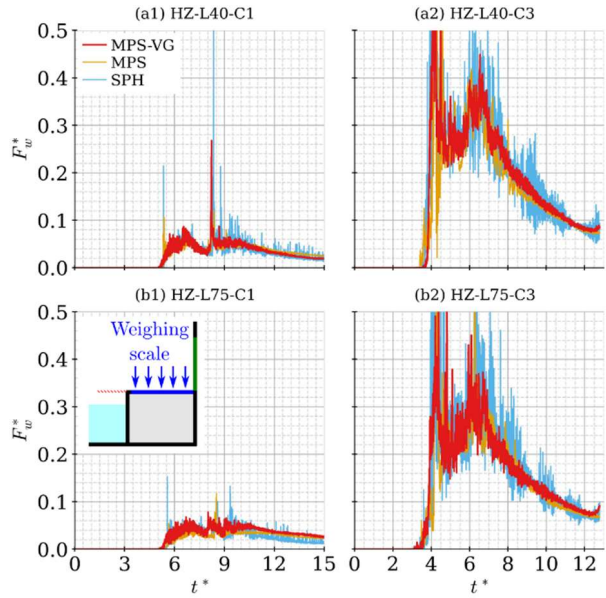

**Figure 13.** Vertical forces on the weighing scale of (a1) HZ-L40-C1, (a2) HZ-L40-C3, (b1) HZ-L75-C1, and (b2) HZ-L75-C3.

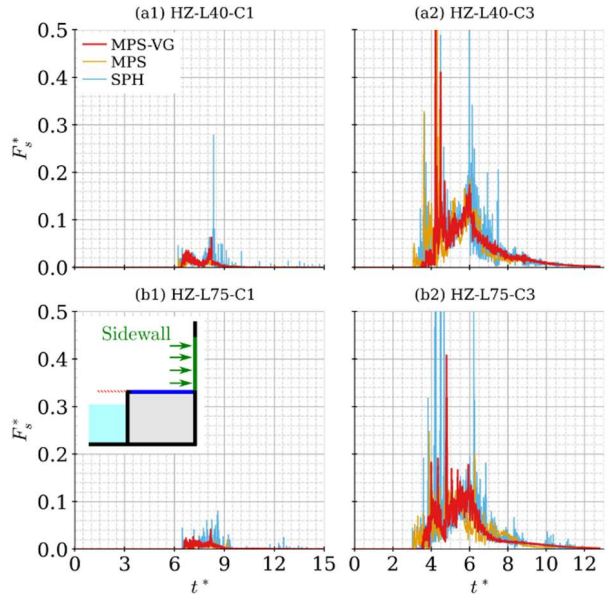

**Figure 14.** Horizontal forces on the sidewall of (a1) HZ-L40-C1, (a2) HZ-L40-C3, (b1) HZ-L75-C1, and (b2) HZ-L75-C3.

### 5.3.2. Impulses on horizontal gratings, weighing scale, and sidewall

Whereas the previous time series depict the main events of the wave runup and impact events, Figure 15 provides quantitative results of the impulse of force on the structures computed over the 0 to 2s time interval. According to the figure, an increase in water column $h_0$ leads to a monotonic increase of dimensionless impulses ($I^*$).

MPS-VG provides intermediate results between MPS and SPH for most situations. The impulses of normal force on the inclined slats computed by MPS-VG are higher than those obtained by MPS (Figure 15(a1) and (a2)), since MPS-VG assumes the ideal situation of complete flow deflection by the inclined slats. Therefore, the loads on the inclined slat computed by MPS-VG are expected to be the upper bound of the actual values.

On the other hand, the impulses obtained by SPH are slightly larger than those provided by MPS-VG, probably due to the pressure oscillations with several large spikes associated with compression waves computed by the weakly compressible SPH, as shown in Figure 12, Figure **13**, and Figure **14**.

Regarding the impulses on the weighing scale and sidewall, MPS-VG, MPS, and SPH obtained very close values, indicating a good agreement for the wave-grating interaction simulations (see Figure 15b and c).

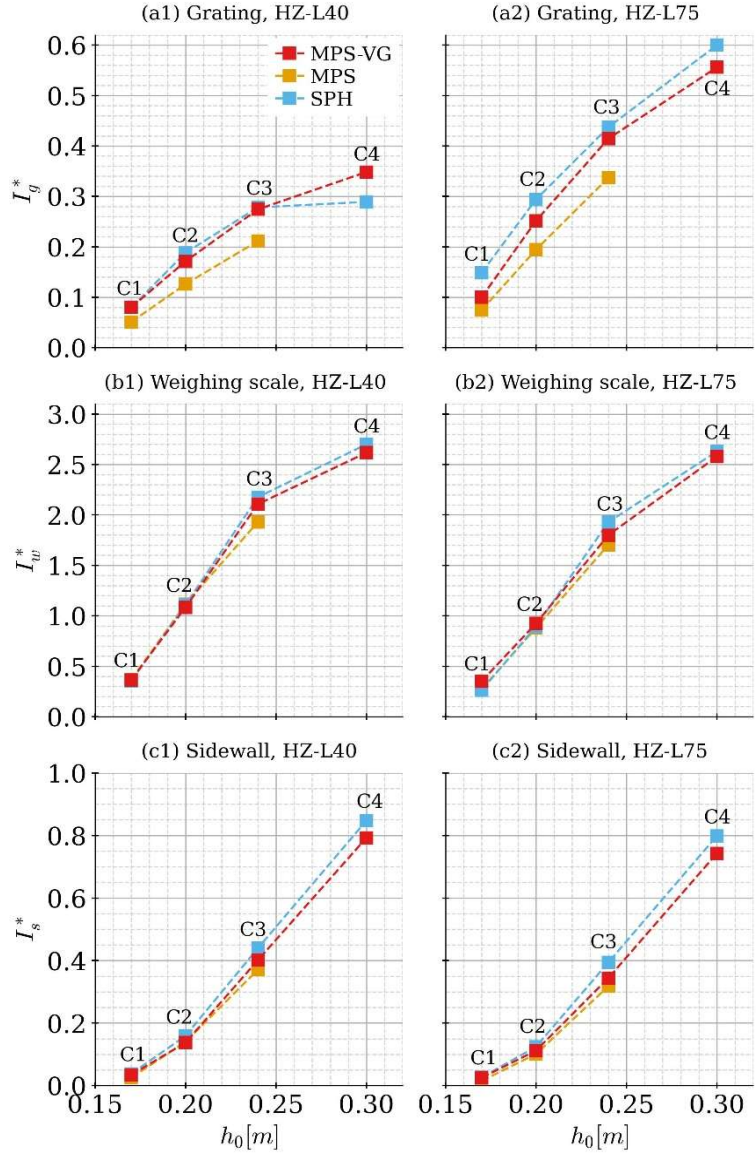

**Figure 15.** Impulses of normal force on the inclined slats of the horizontal gratings (a1) and (a2), vertical force on the weighing scale (b1) and (b2), and horizontal force on the sidewall (c1) and (c2) protected by horizontal gratings of length $l_b = 0.040$m and 0.075m.

### 5.3.3. Water volume above the fixed structure

The average volume of water ($V^*$) during the 2.0 s simulation in region $A_0$ (illustrated in Figure 2) is summarized in Figure 16 for analyses of the green water propagation intensity above the fixed structure. Similar to the trend of the impulses, $V^*$ increases monotonically as $h_0$ increases. The volumes computed by MPS-VG, MPS, and SPH are close, demonstrating the effectiveness of the simplified VG modeling.

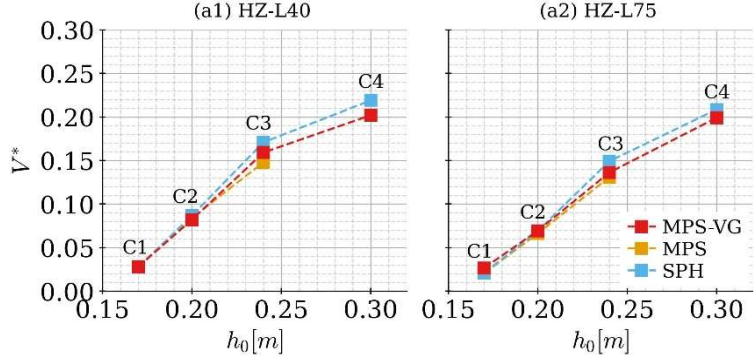

**Figure 16.** Average water volume above the fixed structure protected by horizontal gratings of length $l_b = 0.040$m and $0.075$m.

### 5.3.4. Discrepancies among the results of MPS-VG, MPS, and SPH models

Towards a quantitative analysis, the normalized centered root-mean-square differences ($E_n$) of the force time series computed by MPS-VG were evaluated in relation to those obtained by MPS and SPH:

$$E_{n,A} = \frac{\sqrt{\frac{1}{N}\sum_{i=n}^{N}[(A_i - \overline{A}) - (B_i - \overline{B})]^2}}{\sigma_B}, \tag{42}$$

where $A$ denotes the forces obtained by MPS-VG, $B$ refers to the forces from MPS or SPH, and $N$ is the number of samples. Mean values are denoted by an overhead bar and standard deviation $\sigma$ is obtained by:

$$\sigma_B = \sqrt{\frac{1}{N}\sum_{i=n}^{N}(B_i - \overline{B})^2}. \tag{43}$$

The normalized centered root-mean-square difference (see Eq. (42)) of the normal forces on the slats of the horizontal grating ($E_{n,g}$), vertical force on the weighing scale ($E_{n,w}$), and horizontal force on the sidewall ($E_{n,s}$) protected by the gratings of length $l_b = 0.040$m and $0.075$m are summarized in Table 7 and Table **8**, respectively. The results suggest a better agreement between the forces on weighing scales and sidewalls than the forces on the gratings since the corresponding values of $E_{n,w}$ and $E_{n,s}$ are lower than $E_{n,g}$, with few exceptions. A reason for such a behavior might be the lower flow deflection angle computed by the MPS simulations due to the relatively small solidity ratio of the grating, as shown in Figure 10(a) and discussed at the beginning of Section 5.1. Moreover, the large high-frequency oscillations (see Appendix B) of the forces on the slats computed by MPS-VG through Eq. (30) may also have contributed to that difference.

**Table 7.** Centered root-mean-square difference ($E_n$) of MPS-VG results for the horizontal gratings of length $l_b = 0.040$m compared to MPS and SPH. Values related to the normal force on the inclined slats of the horizontal grating ($E_{n,g}$), vertical force on the weighing scale ($E_{n,w}$), and horizontal force on the sidewall ($E_{n,s}$).

| Condition | Root-mean-square differences between MPS-VG and MPS | | | Root-mean-square differences between MPS-VG and SPH | | |
|---|---|---|---|---|---|---|
| $l_b = 0.040m$ | $E_{n,g}$ (horizontal grating) | $E_{n,w}$ (weighing scale) | $E_{n,s}$ (side wall) | $E_{n,g}$ (horizontal grating) | $E_{n,w}$ (weighing scale) | $E_{n,s}$ (side wall) |
| C1 | 1.15 | 0.50 | 1.02 | 0.90 | 0.69 | 0.80 |
| C2 | 1.14 | 0.43 | 0.71 | 1.12 | 0.45 | 0.74 |
| C3 | 1.02 | 0.69 | 1.32 | 1.02 | 0.62 | 0.90 |
| C4 | - | - | - | 1.07 | 0.62 | 0.70 |

**Table 8.** Centered root-mean-square difference ($E_n$) of MPS-VG results for the horizontal gratings of length $l_b = 0.075$m compared to MPS and SPH. Values related to the normal force on the inclined slats of the horizontal grating ($E_{n,g}$), vertical force on the weighing scale ($E_{n,w}$), and horizontal force on the sidewall ($E_{n,s}$).

| Condition | Root-mean-square differences between MPS-VG and MPS | | | Root-mean-square differences between MPS-VG and SPH | | |
|---|---|---|---|---|---|---|
| $l_b = 0.075$ | $E_{n,g}$ (horizontal grating) | $E_{n,w}$ (weighing scale) | $E_{n,s}$ (side wall) | $E_{n,g}$ (horizontal grating) | $E_{n,w}$ (weighing scale) | $E_{n,s}$ (side wall) |
| C1 | 0.90 | 0.49 | 1.17 | 0.65 | 0.59 | 0.78 |
| C2 | 0.91 | 0.49 | 0.69 | 0.65 | 0.59 | 0.69 |
| C3 | 0.88 | 0.49 | 0.81 | 0.72 | 0.63 | 0.87 |
| C4 | - | - | - | 0.78 | 0.71 | 0.71 |

The relative discrepancies ($\varepsilon$) between the impulses or volumes computed by MPS-VG and those computed by MPS and SPH are calculated as

$$\varepsilon = \frac{\chi_{MPS-VG} - \chi_{REF}}{\chi_{REF}}, \quad (44)$$

where $\chi$ denotes dimensionless impulses $I^*$ or volumes $V^*$ computed by MPS-VG ($\chi_{MPS-VG}$) and subscript *REF* refers to the results of MPS or SPH.

Table 9 and Table **10** summarize the relative discrepancies of the impulses of the normal force on the inclined slats of the horizontal grating ($\varepsilon_g$), vertical force on the weighing scale ($\varepsilon_w$), horizontal force on the sidewall ($\varepsilon_s$), and average water volume ($\varepsilon_V$) above the fixed structure protected by the horizontal gratings of lengths $l_b = 0.040m$ and $l_b = 0.075$m, respectively.

Overall, most of the relative discrepancies for impulses and volumes are below $\varepsilon = 10\%$, indicating good agreements between the results of MPS-VG and those from MPS and SPH. The relative discrepancies of the impulses on the inclined slats tend to be larger than those on the weighing scale and sidewall due to the lower momentum transfer associated with a smaller flow deflection angle computed by the MPS simulations caused by the relatively small solidity ratio of the grating. Moreover, as a tendency in relation to the MPS results, larger relative discrepancies of the impulses on the slats occur in the shorter horizontal grating and under calmer conditions (C1), since the values of impulse are small under C1 and even minimal changes in the computed values result in significant relative discrepancies.

Table 9. Relative discrepancies ($\varepsilon$) of MPS-VG results for the horizontal gratings of length $l_b = 0.040m$ compared to MPS and SPH. Discrepancies of the impulses of normal forces on the inclined slats of the horizontal grating ($\varepsilon_g$), vertical force on the weighing scale ($\varepsilon_w$), horizontal force on the sidewall ($\varepsilon_s$), and average water volume ($\varepsilon_V$) above the fixed structure - values in percentage.

| Condition | Relative discrepancies between MPS-VG and MPS | | | | Relative discrepancies between MPS-VG and SPH | | | |
|---|---|---|---|---|---|---|---|---|
| $l_b = 0.040m$ | $\varepsilon_g$ (%) (horizontal grating) | $\varepsilon_w$ (%) (weighing scale) | $\varepsilon_s$ (%) (side wall) | $\varepsilon_V$ (%) (volume) | $\varepsilon_g$ (%) (horizontal grating) | $\varepsilon_w$ (%) (weighing scale) | $\varepsilon_s$ (%) (side wall) | $\varepsilon_V$ (%) (volume) |
| C1 | 57.3 | 0.5 | 33.0 | 1.2 | 0.7 | 3.2 | 9.7 | 1.1 |
| C2 | 35.8 | 2.9 | 1.5 | 2.6 | 9.1 | 2.1 | 13.6 | 5.4 |
| C3 | 29.7 | 9.2 | 8.2 | 7.6 | 1.4 | 3.2 | 8.7 | 6.9 |
| C4 | - | - | - | - | 20.5 | 3.1 | 6.5 | 7.8 |

Table 10. Relative discrepancies ($\varepsilon$) of MPS-VG results for the horizontal gratings of length $l_b = 0.075$m compared to MPS and SPH. Discrepancies of the impulses of normal forces on the inclined slats of the horizontal grating ($\varepsilon_g$), vertical force on the weighing scale ($\varepsilon_w$), horizontal force on the sidewall ($\varepsilon_s$), and average water volume ($\varepsilon_V$) above the fixed structure. Values in percentage.

| Condition | Relative discrepancies between MPS-VG and MPS | | | | Relative discrepancies between MPS-VG and SPH | | | |
|---|---|---|---|---|---|---|---|---|
| $l_b = 0.075m$ | $\varepsilon_g$ (%) (horizontal grating) | $\varepsilon_w$ (%) (weighing scale) | $\varepsilon_s$ (%) (side wall) | $\varepsilon_V$ (%) (volume) | $\varepsilon_g$ (%) (horizontal grating) | $\varepsilon_w$ (%) (weighing scale) | $\varepsilon_s$ (%) (side wall) | $\varepsilon_V$ (%) (volume) |
| C1 | 33.9 | 28.7 | 67.5 | 29.7 | 32.7 | 33.6 | 4.1 | 27.4 |
| C2 | 29.4 | 5 | 10.7 | 4.6 | 14.4 | 3.4 | 9.9 | 0.1 |
| C3 | 22.9 | 5.7 | 8.3 | 4.1 | 5.2 | 6.7 | 12.7 | 8.7 |
| C4 | - | - | - | - | 7.3 | 1.9 | 7 | 4.5 |

### 5.4. Vertical grating

Figure 17 and Figure **18** show the snapshots of the wave impact against the vertically placed grating of length $l_b = 0.075$m for condition C3 (VT-L75-C2) obtained by MPS and MPS-VG, respectively. The color scale of the fluid particles is related to their pressure.

At $t^* = 3.196$, the grating slats deflect the plunging wave and, subsequently, at $t^* = 4.539$, a considerable amount of water passes through the grating, flows over the weighing scale, and impacts the downstream sidewall.

In general, the wave profile and pressure field computed by MPS and MPS-VG are in good agreement. In the MPS simulations, the plunging wave is divided into jet flows by the slats of the vertical grating and the flow deflection angle is smaller than the slats inclination, as observed in the cases with horizontal gratings. In contrast, MPS-VG obtained a single jet flow, which closely reflects the ideal situation with complete deflection of the flow by very thin grating slats. Moreover, the abrupt change of the $pnd_i$ of fluid particles deflected by VG leads to a sharp pressure gradient near the VG region, as illustrated in Figure 18(a). Some differences between the flow behavior computed by MPS and MPS-VG after $t^* = 4.507$ (Figure 17(b)-(c) and Figure **18** (b)-(c)) are expected due to the chaotic nature of the breaking wave (see Wei et al. (2018)).

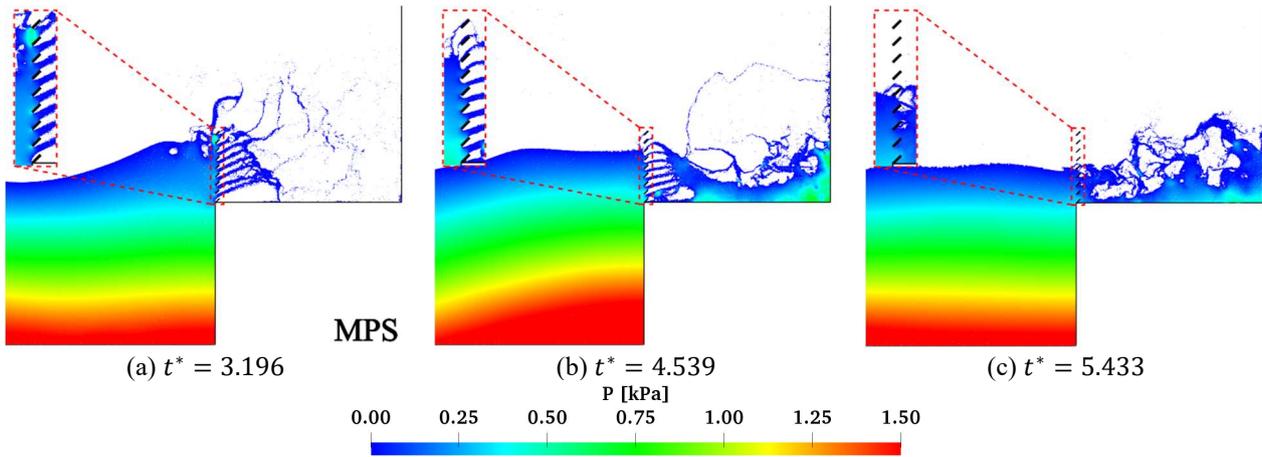

(a) $t^* = 3.196$     (b) $t^* = 4.539$     (c) $t^* = 5.433$

**Figure 17.** Snapshots of wet dam breaking event on a fixed structure for VT-L75-C3 at three instants obtained by MPS simulations. The color scale represents pressure magnitude.

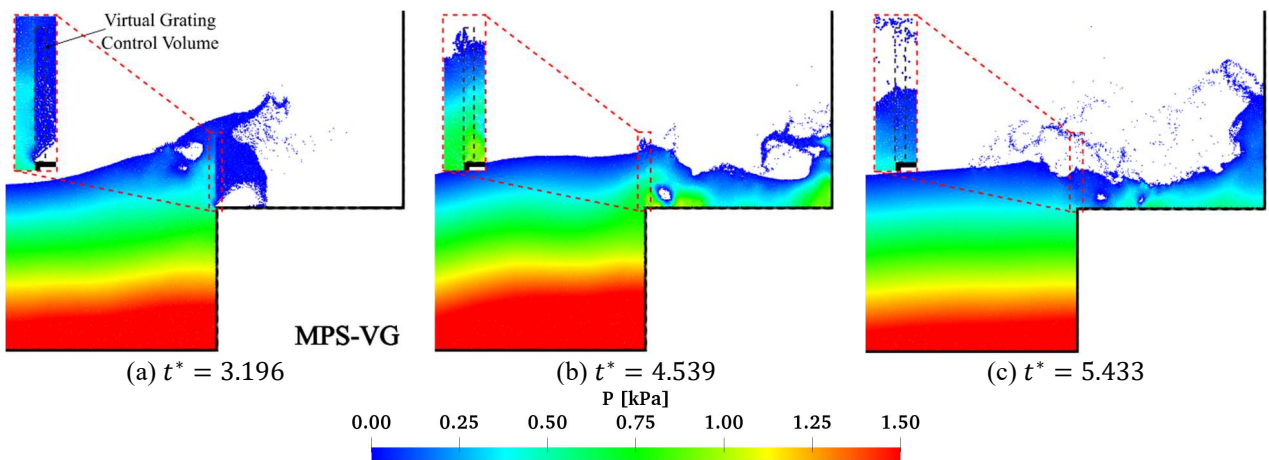

(a) $t^* = 3.196$     (b) $t^* = 4.539$     (c) $t^* = 5.433$

**Figure 18.** Snapshots of wet dam breaking event on a fixed structure for VT-L75-C3 at three instants obtained by MPS-VG simulations. The color scale represents pressure magnitude.

### 5.4.1. Forces on vertical gratings, weighing scale, and sidewall

Figure 19 depicts the time series of the normal forces on the inclined slats of vertical gratings of lengths $l_b = 0.040$m and 0.075m for calm (C1) and rough (C3) conditions. As detailed in Appendix B, the forces computed by MPS-VG are filtered by a low pass filter with a 100 Hz cutoff frequency.

Under C1, a force peak occurs at $t^* \sim 5$ when the wave impacts the vertical grating. The force increases again after $t^* \sim 9$ due to the return of the flow on the fixed structure. Under C3, a violent wave impact on the grating occurs at $t^* \sim 2.5$, followed by another impact near $t^* \sim 7$ to 8 due to the collapse of the incoming wave on the fixed structure.

Figure 20 shows the time series of the vertical forces on the weighing scale protected by vertical gratings of lengths $l_b = 0.040$m and 0.075m for conditions C1 and C3. According to Figure 20(a1) and (b1), all computed forces for C1 exhibit a similar sequence of events. The force increases at $t^* \sim 4.5$ when the flow passes through the vertical grating and, after the runup along the sidewall, part of the fluid falls under gravity onto the weighing scale and a sharp force peak occurs at $t^*\sim 8$. The force then decreases progressively, with the flow returning to the fixed structure. Concerning relatively severe condition C3 (Figure 20(a2) and (b2)), a large amount of water associated with the breaking of the incoming wave leads to the first peak impact force at $t^*\sim 4$ followed by an abrupt decay. Subsequently, the force increases rapidly at $t^*\sim 6.5$ due to the impact of the fluid that falls after the runup on the downstream sidewall and then decays until $t^*\sim 12$.

Figure 21 shows the horizontal forces computed on the sidewall protected by vertical gratings. Whereas the force magnitude is very small for C1 (Figure 21(a1) and (b1)), for C3, tow force peaks occur on the sidewall near $t^*\sim 4.0$ and $t^*\sim 6.0$ during the buildup and collapse of the runup, respectively, as illustrated in Figure 21(a2) and (b2).

Overall, the time series of the normal forces on the inclined slats of the vertical gratings computed by the low-resolution MPS-VG and high-resolution MPS and SPH are in reasonable agreement. Therefore, the simplified VG modeling reproduced all the relevant events of the wave impact on the gratings. The values computed by MPS-VG tend to be higher than those obtained by MPS. As addressed in Section 5.3.2, this is expected, since the ideal condition of complete flow deflection due to the inclined slats is assumed in MPS-VG. Consequently, the forces acting on the inclined slat computed by MPS-VG are the upper bound of the actual values. On the other hand, those obtained by SPH are close to MPS-VG despite the raw data of SPH showing a significant pressure oscillation with several large spikes due to compression waves computed by the weakly compressible algorithm.

MPS-VG, MPS, and SPH yielded close results concerning the forces on the weighing scale and sidewall protected by the vertical gratings.

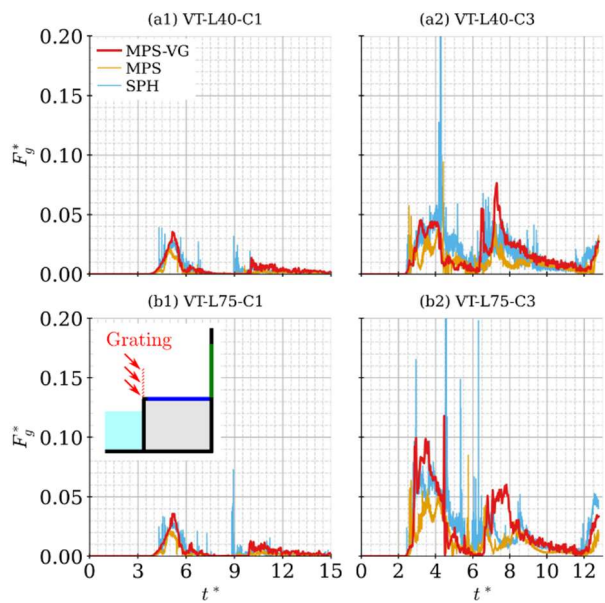

**Figure 19.** Normal forces on the inclined slats of (a1) VT-L40-C1, (a2) VT-L40-C3, (b1) VT-L75-C1, and (b2) VT-L75-C3. MPS-VG force filtered by a low-pass filter ($f_c = 100 Hz$).

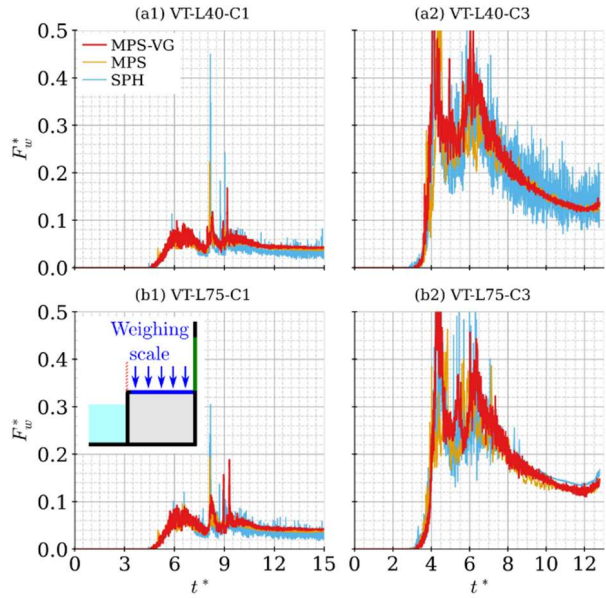

**Figure 20.** Vertical forces on the weighing scale of (a1) VT-L40-C1, (a2) VT-L40-C3, (b1) VT-L75-C1, and (b2) VT-L75-C3.

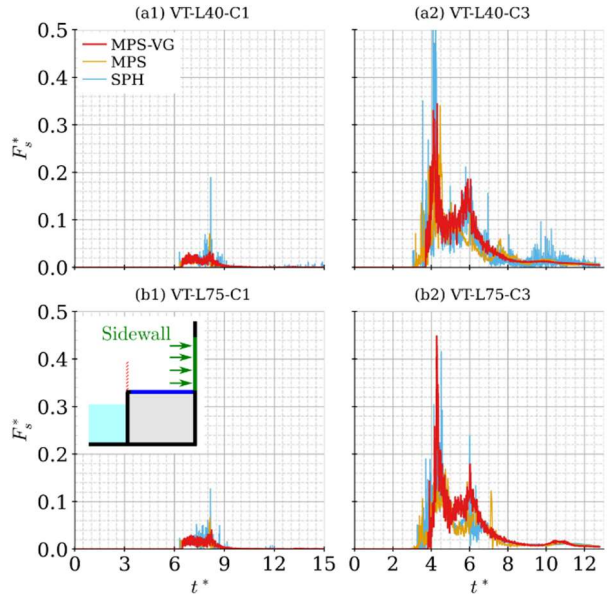

**Figure 21.** Horizontal forces on the sidewall of (a1) VT-L40-C1, (a2) VT-L40-C3, (b1) VT-L75-C1, and (b2) VT-L75-C3.

### 5.4.2. Impulses on vertical gratings, weighing scale, and sidewall

Figure 22 shows the dimensionless impulses of the normal force on the inclined slats of vertical grating $I_g^*$, weighing scale $I_w^*$, and sidewall $I_s^*$ computed over the 0 to 2s time interval. As the upstream column height $h_0$ increases, the dimensionless impulses also increase $I^*$.

According to Figure 22(a1) and (a2), the impulses on the inclined slats of the vertical gratings computed by MPS-VG are higher than those obtained by MPS and SPH, except for condition C4 and $l_b = 0.040$ m grating length (see Figure 22(a1)). Again, this is because the inclined slats completely deflect all the fluid particles in VG, whereas in MPS, the angle of the jet flows is smaller than the inclination of the slats, which is associated with a lower momentum transfer from the fluid to the slats. Moreover, the impulses obtained by SPH are close to the MPS-VG ones, probably due to the larger oscillations of the force associated with compression waves computed by SPH. All impulses computed by MPS-VG, MPS, and SPH are very close for the weighing scale and sidewall, indicating a good reproduction of the wave-grating interaction by VG (see Figure 22(b) and (c)).

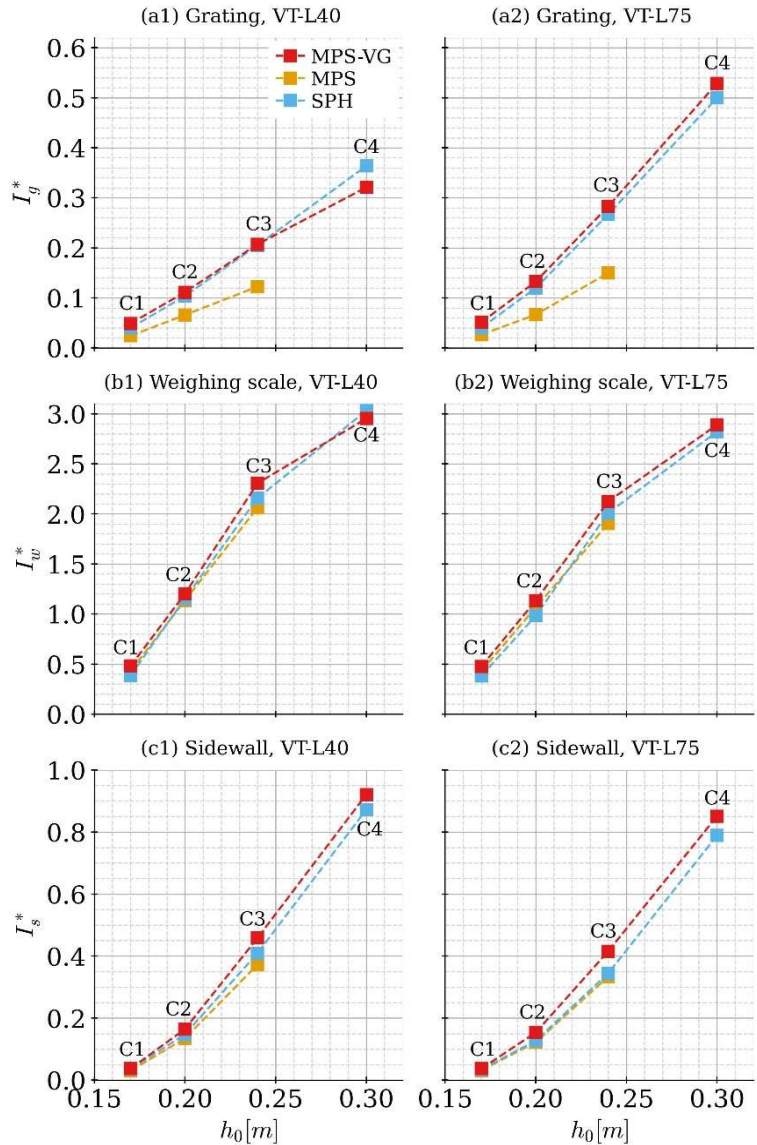

**Figure 22.** Impulses of normal force on the inclined slats of vertical gratings (a1) and (a2), vertical force on the weighing scale (b1), and (b2) horizontal force on the sidewall (c1) and (c2) protected by vertical gratings of length $l_b = $ 0.040m and 0.075m.

### 5.4.3. Water volume on the fixed structure

Figure 23 depicts the average volume of water ($V^*$) in region $A_0$ (see Figure 2) above the fixed structure protected by vertical gratings during the 2.0 s simulation. Overall, the average volume of water on the fixed structure increases as water column $h_0$ increases and the results from MPS-VG, MPS, and SPH are close for all conditions.

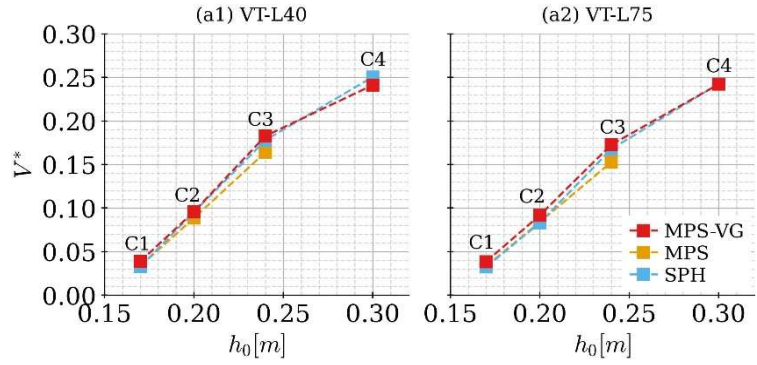

**Figure 23.** Average water volume above the fixed structure protected by vertical gratings of length $l_b = 0.040$m and $0.075$m.

### 5.4.4. Discrepancies among the results of MPS-VG, MPS, and SPH models

Table 11 and Table 12 show the centered root-mean-square differences (see Eq. (42)) between the force time series computed by MPS-VG and MPS for the vertical gratings of lengths $l_b = 0.040$m and $0.075$m, respectively. In general, the differences in the inclined slats of grating $E_{n,g}$ are larger than those in weighing scale $E_{n,w}$ and sidewall $E_{n,s}$, showing the same tendency observed in the horizontal grating.

**Table 11.** Centered root-mean-square difference ($E_n$) of MPS-VG results for the vertical gratings of length $l_b = 0.040$m compared to MPS and SPH. Values related to the forces on the inclined slats of vertical grating ($E_{n,g}$), weighing scale ($E_{n,w}$), and sidewall ($E_{n,s}$).

| Condition | Root-mean-square differences between MPS-VG and MPS | | | Root-mean-square differences between MPS-VG and SPH | | |
|---|---|---|---|---|---|---|
| $l_b = 0.040m$ | $E_{n,g}$ (vertical grating) | $E_{n,w}$ (weighing scale) | $E_{n,s}$ (side wall) | $E_{n,g}$ (vertical grating) | $E_{n,w}$ (weighing scale) | $E_{n,s}$ (side wall) |
| C1 | 1.31 | 0.43 | 0.65 | 0.82 | 0.64 | 0.72 |
| C2 | 1.34 | 0.39 | 0.77 | 0.95 | 0.42 | 0.70 |
| C3 | 1.64 | 0.64 | 0.82 | 1.06 | 0.46 | 0.64 |
| C4 | - | - | - | 1.13 | 0.51 | 0.67 |

**Table 12.** Centered root-mean-square difference ($E_n$) of MPS-VG results for the vertical gratings of length $l_b = 0.075$m compared to MPS and SPH. Values related to the forces on the inclined slats of vertical grating ($E_{n,g}$), weighing scale ($E_{n,w}$), and sidewall ($E_{n,s}$).

| Condition | Root-mean-square differences between MPS-VG and MPS | | | Root-mean-square differences between MPS-VG and SPH | | |
|---|---|---|---|---|---|---|
| $l_b = 0.075m$ | $E_{n,g}$ (vertical grating) | $E_{n,w}$ (weighing scale) | $E_{n,s}$ (side wall) | $E_{n,g}$ (vertical grating) | $E_{n,w}$ (weighing scale) | $E_{n,s}$ (side wall) |
| C1 | 1.22 | 0.45 | 0.71 | 0.81 | 0.61 | 0.7 |
| C2 | 1.51 | 0.39 | 0.82 | 0.8 | 0.47 | 0.79 |
| C3 | 1.70 | 0.62 | 0.91 | 0.97 | 0.51 | 0.81 |
| C4 | - | - | - | 1.07 | 0.68 | 1.00 |

The relative discrepancies ($\varepsilon$) (see Eq. (44)) for the impulses of forces on the inclined slats ($\varepsilon_g$), weighing scale ($\varepsilon_w$), sidewall ($\varepsilon_s$), and average water volume ($\varepsilon_V$) above the fixed structure considering the vertical gratings of lengths $l_b = 0.040m$ and $l_b = 0.075$m are provided in Table 13 and Table **14**, respectively.

The agreement between the impulses of forces on the weighing scale and sidewall is satisfactory, with relative discrepancies $\varepsilon_w \leq 25\%$ for the weighing scale and $\varepsilon_s \leq 25.6\%$ for the sidewall. The relative discrepancies for volumes $\varepsilon_V \leq 19.1\%$ also indicate a reasonable agreement among the results of MPS-VG, MPS and SPH. Concerning the impulses on the inclined slats of the vertical gratings, the relatively large discrepancies between MPS-VG and MPS reinforce the feature of VG in providing the upper limit of the force on slats, as discussed in Section 5.4.2. On the other hand, the impulses on the inclined slats of the vertical gratings calculated by MPS-VG are in good agreement with the SPH results.

**Table 13.** Relative discrepancies ($\varepsilon$) of MPS-VG results for the vertical gratings of length $l_b = 0.040$m compared to MPS and SPH. Discrepancies of the impulses of normal forces on the inclined slats of the vertical grating ($\varepsilon_g$), vertical force on the weighing scale ($\varepsilon_w$), vertical force on the sidewall ($\varepsilon_s$), and average water volume ($\varepsilon_V$) above the fixed structure. Values in percentage.

| Condition | Relative discrepancies between MPS-VG and MPS | | | | Relative discrepancies between MPS-VG and SPH | | | |
|---|---|---|---|---|---|---|---|---|
| $l_b = 0.040m$ | $\varepsilon_g$ (%) (vertical grating) | $\varepsilon_w$ (%) (weighing scale) | $\varepsilon_s$ (%) (side wall) | $\varepsilon_V$ (%) (volume) | $\varepsilon_g$ (%) (vertical grating) | $\varepsilon_w$ (%) (weighing scale) | $\varepsilon_s$ (%) (side wall) | $\varepsilon_V$ (%) (volume) |
| C1 | 101.3 | 11.6 | 20.9 | 15.7 | 22.6 | 24.9 | 3.2 | 18.7 |
| C2 | 68.8 | 5.9 | 22.2 | 8.3 | 7.2 | 4.1 | 11.6 | 0.5 |
| C3 | 69.2 | 11.9 | 23 | 11.4 | 0.9 | 6.9 | 11.9 | 2.7 |
| C4 | - | - | - | - | 11.7 | 2.6 | 5.6 | 3.8 |

**Table 14.** Relative discrepancies ($\varepsilon$) of MPS-VG results for the vertical gratings of length $l_b = 0.075$m compared to MPS and SPH. Discrepancies of the impulses of forces on the inclined slats of the vertical grating ($\varepsilon_g$), vertical force on the weighing scale ($\varepsilon_w$), horizontal force on the sidewall ($\varepsilon_s$), and average water volume ($\varepsilon_V$) above the fixed structure. Values in percentage.

| Condition | Relative discrepancies between MPS-VG and MPS | | | | Relative discrepancies between MPS-VG and SPH | | | |
|---|---|---|---|---|---|---|---|---|
| $l_b = 0.075m$ | $\varepsilon_g$ (%) (vertical grating) | $\varepsilon_w$ (%) (weighing scale) | $\varepsilon_s$ (%) (side wall) | $\varepsilon_V$ (%) (volume) | $\varepsilon_g$ (%) (vertical grating) | $\varepsilon_w$ (%) (weighing scale) | $\varepsilon_s$ (%) (side wall) | $\varepsilon_V$ (%) (volume) |
| C1 | 92.8 | 10.9 | 17.9 | 15.5 | 26.8 | 25 | 7.6 | 19.1 |
| C2 | 98.7 | 6.3 | 25.6 | 9.3 | 10.6 | 15.2 | 20.5 | 10.6 |
| C3 | 88.8 | 11.6 | 24.6 | 13.1 | 5.9 | 5.8 | 20.6 | 3 |
| C4 | - | - | - | - | 5.8 | 2.4 | 7.9 | 0.2 |

### 5.5. Computational performance of MPS-VG

Since the fully particle-based MPS and proposed MPS-VG models adopt the same semi-implicit algorithm, the MPS simulations were used as a reference for the analysis of the processing times. The vertical gratings of length $l_b = 0.075$m and conditions C1, C2, or C3 were considered. The speedup represents the ratio between the processing times of MPS and MPS-VG.

As shown in Table 15, MPS-VG drastically reduced the processing times in all conditions since it enables the adoption of a relatively low particle model resolution. Notably, MPS-VG provides speedups between 90 and 115 times compared to fully particle-based MPS models. Another benefit of MPS-VG is the low memory consumption since it requires fewer particles in comparison to the high-resolution MPS models.

Table 15. Processing time and number of particles of MPS and MPS-VG modeling for simulations of 2.0s.

| Condition | MPS | | MPS-VG | | Speedup |
|---|---|---|---|---|---|
| | Num. of particles | Processing time* [h] | Num. of particles | Processing time* [h] | |
| C1 | 2 069 340 | 222.47 | 133 330 | 2.45 | **90.8x** |
| C2 | 2 220 300 | 271.82 | 143 250 | 2.68 | **101.4x** |
| C3 | 2 439 810 | 355.23 | 157 000 | 3.10 | **114.6x** |

*Intel® Xeon® Processor E5-2680 v2 2.80GHz, 10 Cores (20 Threads)

Finally, huge speedups were obtained for 2D models. Further savings in processing time and memory consumption are expected for 3D simulations.

## 6. Wave-breaking performance of inclined-slats gratings

This section addresses the performance of gratings with the inclined slats in mitigating hydrodynamic loads. The relative differences between the impulses on the weighing scale and sidewall and the green water volumes computed by simulations with and without gratings are considered. The relative differences were calculated as

$$X_r = \frac{X - X_{ref}}{X_{ref}}, \tag{45}$$

where $X$ denotes the values with grating, which were computed by MPS-VG, and $X_{ref}$ represents those without the grating, computed by MPS.

The effects of position (horizontal or vertical) and length ($l_b = 0.040$ m or $0.075$ m) of the gratings on the protection of the fixed structure are summarized in Figure 24. According to the figure, the percentage reduction of the hydrodynamic loads achieved by the gratings decreases as the initial water column increases from C1 to C4, except for the vertical grating under C1. The horizontal gratings (light and dark red bars) provide better protection than the vertical ones (light and dark blue bars) in mitigating the impulses for all conditions (see Figure 24(a) and (b)) and significantly reduce the average green water volume (Figure 24(c)). The superior performance of the gratings placed horizontally can be explained by the considerable reduction in the wave runup along the freeboard. Nevertheless, the effects of increasing the grating length are almost negligible for the less severe conditions, especially for the vertical gratings.

The gratings become more effective as length increases, leading to more efficient breakwaters, especially for horizontal gratings. Expressive reductions of impulses or volumes are obtained for horizontal gratings of length $l_b = 0.075$m for all conditions (dark red bars in Figure 24). Nonetheless, increasing length leads to a relatively small gain of protection to less severe conditions.

According to Figure 24(a), positive values of the relative difference of impulse on the weighing scale were obtained for the vertical gratings in C4 and are associated with fluid accumulation on the fixed structure due to the blockage of the backflow by the vertical grating. Similarly, a degradation of vertical grating performance on the water above the deck (average green water volume) occurs in C3 and C4, as shown in Figure 24(c).

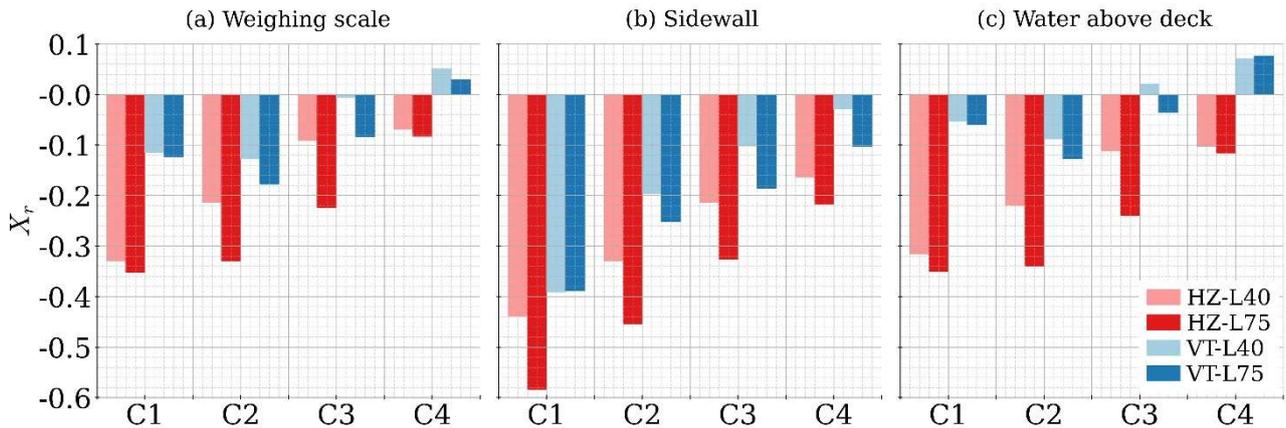

**Figure 24.** Effects of horizontal (HZ) and vertical (VT) gratings of lengths $l_b = 0.040$m and 0.075m on the impulses of force on the weighing scale (a), sidewall (b), and average green water volume (c) for C1 to C4. A negative value with a greater magnitude means better performance compared to the reference case without grating.

## 7. Conclusions

This study has proposed a Virtual Grating (VG) model for an efficient modeling of the multi-scale problem of wave impact on gratings with inclined slats. VG was incorporated in MPS, here called MPS-VG, by replacing the conventional particle-based wall modeling of the inclined slats. Its capabilities and drawbacks were evaluated by numerical simulations of wave-structure interaction problems (WSI).

Overall, the simulations showed the predicted jet flow in MPS-VG is qualitatively closer to the one expected in ideal conditions involving an array of thin slats with a large solidity ratio. Moreover, the wave profile and pressure field calculated by MPS-VG and MPS matched. Notwithstanding, some discrepancies between the flow profiles obtained by MPS-VG and MPS occurred during the incoming wave-breaking process, which is reasonable due to the chaotic nature at this stage.

According to the quantitative analysis, wave impact loads on the gratings placed horizontally and vertically, and water volumes computed by MPS-VG agreed with those obtained by MPS and SPH. Despite spurious numerical oscillations, the time series of all forces computed by the low-resolution MPS-VG and high-resolution MPS and SPH models are in good agreement. Therefore, VG can reproduce all relevant events of the wave impact on the grating with inclined slats.

MPS-VG provides values of hydrodynamic impulses on the horizontal gratings between those obtained by MPS and SPH, thus showing satisfactory performance. The hydrodynamic impulses on vertical gratings computed by MPS-VG tend to be slightly larger than the ones of MPS and SPH. This is reasonable since, in VG, all the fluid particles inside the grating are completely deflected by the inclined slats and the impulse obtained by MPS-VG is expected to be the upper bound of the actual values.

A great advantage of the proposed modeling is its computational efficiency. Whereas MPS requires very high resolutions to adequately reproduce the influences of tiny grating slats on the wave-grating interaction, MPS-VG enables the adoption of a medium-resolution model. As a result, MPS-VG provides speedups of processing time between 90 and 115 times compared to fully particle-based MPS models. Another benefit of MPS-VG is its low memory consumption since it requires fewer particles in comparison to high-resolution MPS models.

Finally, the effects of the grating structures as wave energy dampers to reduce the wave loads and water volume on the fixed structure were evaluated, revealing:

- Grating position: horizontal gratings perform better than vertical ones by mitigating the wave runup along the freeboard and avoiding a large fluid accumulation on the fixed structure during backflow;
- Grating dimension: an increase in the grating length leads to more effective breakwaters, especially for horizontal gratings, despite its relatively small contribution in less severe conditions; and
- Initial water column: the percentual of hydrodynamic load reduction provided by the gratings becomes smaller for more severe conditions (from C1 to C4).

In summary, the results indicate MPS-VG can be a valuable CFD tool for the preliminary design, construction, and optimization of grating-type breakwaters with inclined slats, which can have a wider variety of practical coastal and offshore engineering applications.

## 8. Future work

Despite the advantages of VG in dealing with complex interactions between waves and gratings formed by inclined slats, it also has some limitations.

Regarding physical modeling, since VG assumes the inclined slat has an infinitesimal thickness, the effects of the slat transverse section and surfaces are neglected. Among such effects, the drag due to the flow through the inclined-slat gratings can be considered by introducing an energy dissipation term, following the ideas adopted by existing simplified models of perforated walls, slatted screens, or porous media.

Moreover, whereas VG deflects the flow completely following the inclination angle of the slats, the effective flow deflection angle tends to be smaller in actual situations. Among several factors, the effective flow deflection angle depends on the grating solidity ratio, which should also be considered in the numerical model. Furthermore, the effects of turbulence and air phase can also be interesting topics to be considered, although such features are still an open field in the context of particle methods (Vacondio, et al., 2020).

The results of numerical performance are in good agreement with those of traditional fully particle-based modeling of MPS and SPH methods. However, the unstable nature of the hydrodynamic force computation demands an in-depth investigation of both cause and solution towards mitigating spurious high-frequency time-domain oscillations.

The computational efficiency of VG was demonstrated through 2D models. Aiming at practical applications, the model can be easily extended to 3D, resulting in further processing time and memory savings. Additionally, it requires fewer particles and a parallelized version using graphics processing units (GPUs) architecture will drastically reduce computation time while circumventing GPU memory limitations.

Finally, as a step forward in validating the current technique, experimental tests involving the multi-scale wave impact on gratings with inclined slats should be conducted.

**Appendix A. Calibration of numerical parameter $C_s$**

The so-called speed of propagation of perturbations $C_s$ (Eq. (18)) was calibrated for ensuring numerical stability and minimizing the influence of numerical parameters on the results. Simulations of a tank with hydrostatic water column height $H_F = 0.15$m were conducted considering a wide range of $C_s$ values and both hydrostatic water column height and hydrostatic pressure at the tank bottom were recorded as references for the calibration. The normalized root-mean-square deviation of the hydrostatic water column height was used in the analysis of the stability of the simulations:

$$NRMSD_H = \frac{1}{H_F}\sqrt{\frac{1}{N}\sum_{i=n}^{N}(H_i - H_F)^2}, \qquad (46)$$

where $H_i$ is the computed height of the free-surface particle amid the tank and $N$ is the number of values computed during a time interval. The standard deviation of pressure time series $\sigma_p$ was calculated by Eq. (43).

Figure 25 displays the deviations of water column height and pressure as a function of $C_s$ for the initial distance between particles of $l_0 = 7.5$mm. The optimal choice for $C_s$ is a compromise between minimizing fluid compressibility (minimum deviation of hydrostatic water column height) and pressure oscillation (minimum deviation of hydrostatic pressure). According to the results in Figure 25, the optimal value $C_s = 2.0$ m/s was adopted for the initial distance between particles of $l_0 = 7.5$mm. For the other resolutions, $C_s$ was estimated by relation $C_s = A\sqrt{gl_0}$, with $A = 7.3$ within the range recommended by Cheng et al. (2021).

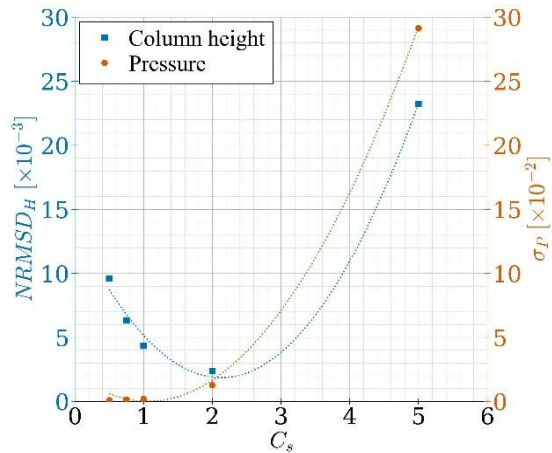

**Figure 25.** Calibration of $C_s$ for MPS considering the initial distance between particles of $l_0 = 7.5$mm.

**Appendix B. Filtering of the forces on the grating**

The filtering process applied to raw data of forces normal to the grating slats computed by MPS-VG are illustrated for conditions VT-L75-C1 and VT-L75-C3, see Table 3. The time series of the two cases are provided in Figure 26(a1) and (b1), respectively and the raw data of the forces computed by MPS-VG based on Eq. (30) are represented in light blue, showing high-frequency oscillations.

Fast Fourier Transform (FFT) was applied to the raw forces time series computed by MPS-VG for checking the relevance of high-frequency components. The force amplitude spectra for VT-L75-C1 and VT-L75-C3 are respectively plotted in Figure 26(a2) and (b2), which also show the presence of small-amplitude high-frequency wave components. Nevertheless, since the energy of each wave component is proportional to the square of its amplitude, the contribution of the high-frequency components is almost negligible.

Therefore, since the main wave components of both calm and severe cases are much lower than 100Hz, a low-pass filter with a $f_c = 100$Hz cut-off frequency was applied to the forces on the gratings obtained by MPS-VG and high-frequency noises were removed. The filtered time series of the forces on grating obtained by MPS-VG are displayed in dark blue in Figure 26(a1) and (b1).

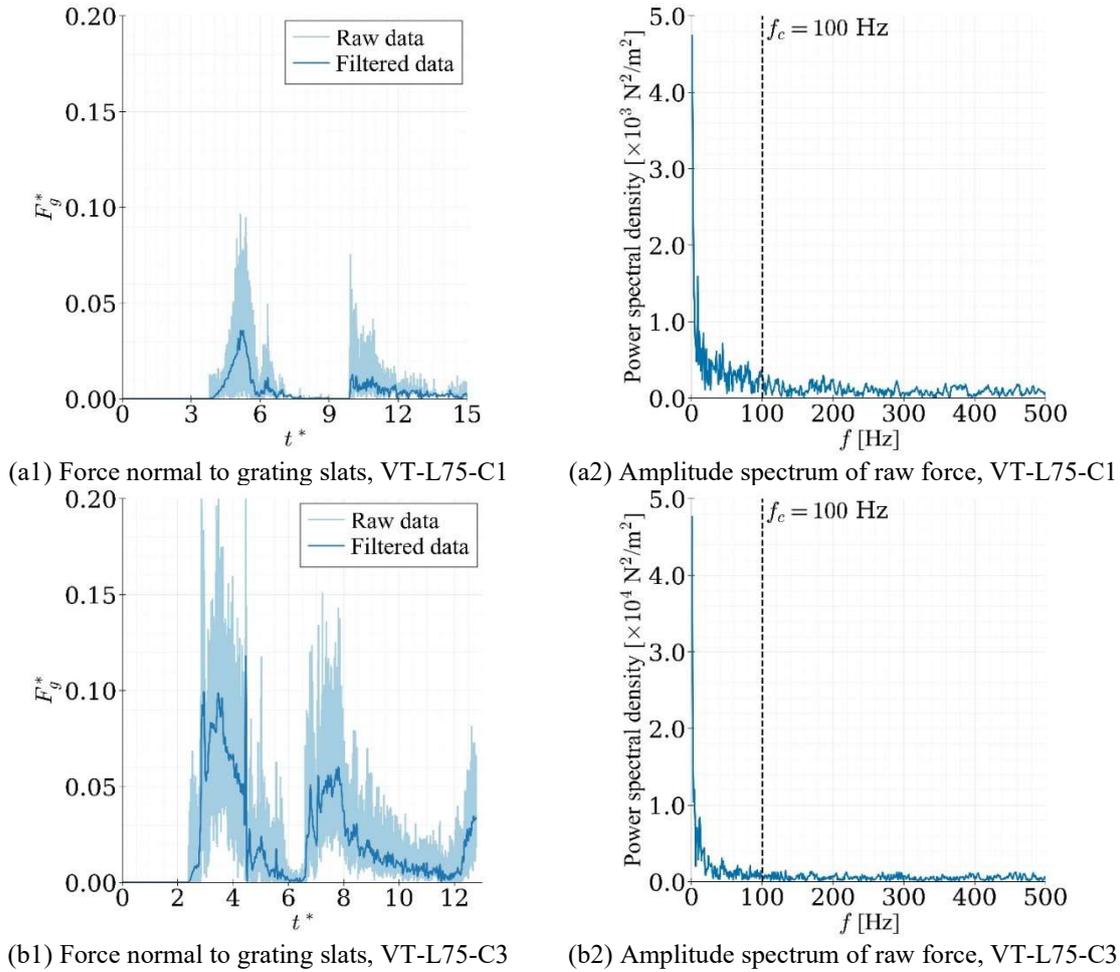

(a1) Force normal to grating slats, VT-L75-C1    (a2) Amplitude spectrum of raw force, VT-L75-C1

(b1) Force normal to grating slats, VT-L75-C3    (b2) Amplitude spectrum of raw force, VT-L75-C3

**Figure 26.** Time series of raw and filtered forces normal to grating slats obtained by MPS-VG for (a1) VT-L75-C1 and (b1) VT-L75-C3. Amplitude spectrum of raw forces data for (a2) VT-L75-C1 and (b2) VT-L75-C3.

**Acknowledgments**

This research was supported by MODEC, Inc., and partially financed by the Coordenação de Aperfeiçoamento de Pessoal de Nível Superior - Brasil (CAPES) - Finance Code 001. R. A. Amaro Jr acknowledges the financial support from the São Paulo Research Foundation FAPESP/CEPID/CeMEAI (grant 2021/11429-4), and F. S. Sousa acknowledges the financial support from CNPq grant 310990/2019-0. The SPH simulations with open-source code DualSPHysics, to which the authors are grateful, were conducted with the use of computational resources of the Center for Mathematical Sciences Applied to Industry (CeMEAI) funded by FAPESP (grant 2013/07375-0). Finally, the authors are particularly grateful to Angela Cristina Pregnolato Giampedro for the correction and improvement of the English text.

**References**

Amaro Jr, R. A., Cheng, L.-Y. & Buruchenko, S. K., 2021. A Comparison Between Weakly-Compressible Smoothed Particle Hydrodynamics (WCSPH) and Moving Particle Semi-Implicit (MPS) Methods for 3D Dam-Break Flows. *International Journal of Computational Methods,* 18(2).

Amaro Jr, R. A., Cheng, L.-Y. & Rosa, S. V., 2019. Numerical study on performance of perforated breakwater for green water. *Journal of Waterway, Port, Coastal, and Ocean Engineering*, 145(6).

Areu-Rangel, O. S. et al., 2021. Green water loads using the wet dam-break method and SPH. *Ocean Engineering,* Volume 219.

Aureli, F., Maranzoni, A., Petaccia, G. & Soares-Frazão, S., 2023. Review of Experimental Investigations of Dam-Break Flows over Fixed Bottom. *Water,* 21 March, Volume 15, p. 1229.

Awad, B. N. & Tait, M. J., 2022. Macroscopic modelling for screens inside a tuned liquid damper using incompressible smoothed particle hydrodynamics. *Ocean Engineering*, Volume 263.

Barcarolo, D. A., Le Touzé, D., Oger, G. & De Vuyst, F., 2014. Adaptive particle refinement and derefinement applied to the smoothed particle hydrodynamics method. *Journal of Computational Physics,* Volume 273, pp. 640-657.

Bellezi, C. A., Cheng, L.-Y., Amaro Jr, R. A. & Tsukamoto, M. M., 2022. Border mapping multi-resolution (BMMR) technique for incompressible projection-based particle methods. *Computer Methods in Applied Mechanics and Engineering*, Volume 396.

Bellezi, C. A., Cheng, L.-Y., Okada, T. & Arai, M., 2019. Optimized perforated bulkhead for sloshing mitigation and control. *Ocean Engineering,* Volume 187.

Buchner, B., 2002. *Green water on ship-type offshore structures,* The Netherlands: Delft University of Technology Delft.

Cheng, L. Y., Amaro Junior, R. A. & Favero, E. H., 2021. Improving stability of moving particle semi-implicit method by source terms based on time-scale correction of particle-level impulses. *Engineering Analysis with Boundary Elements,* Volume 131, pp. 118-145.

Cho, I. H. & Kim, M. H., 2008. Wave absorbing system using inclined perforated plates. *Journal of Fluid Mechanics*, Volume 608, pp. 1-20.

Chorin, A. J., 1967. The numerical solution of the Navier-Stokes equations for an incompressible fluid. *Bulletin of the American Mathematical Society,* 73(6), pp. 928-931.

Courant, R., Friedrichs, K. & Levy, H., 1967. On the partial difference equations of mathematical physics. *IBM Journal of Research and Development,* 11(2), pp. 215-234.

Cummins, S. J. & Rudman, M., 1999. An SPH projection method. *Journal of Computational Physics,* 152(2), pp. 584-607.

Domínguez, J. M. et al., 2022. DualSPHysics: from fluid dynamics to multiphysics problems. *Computational Particle Mechanics,* 1 September, Volume 9, p. 867–895.

Faltinsen, O. M., Firoozkoohi, R. & Timokha, A. N., 2010. Analytical modeling of liquid sloshing in a two-dimensional rectangular tank with a slat screen. *Journal of Enginnering Mathematics*, 04 August, pp. 93-109.

Francis, V. et al., 2023. Solitary wave interaction with upright thin porous barriers. *Ocean Engineering*, Volume 268.

Gao, T., Qiu, H. & Fu, L., 2023. Multi-level adaptive particle refinement method with large refinement scale ratio and new free-surface detection algorithm for complex fluid-structure interaction problems. *Journal of Computational Physics*, Volume 473.

Gingold, R. A. & Monagham, J. J., 1977. Smoothed particle hydrodynamics: theory and application to non-spherical stars. *Monthly Notices of the Royal Astronomical Society,* Volume 181, pp. 375-389.

González-Cao, J. et al., 2019. On the accuracy of DualSPHysics to assess violent collisions with coastal structures. *Computers & Fluids,* 22 January, Volume 179, p. 604–612.

Hernández-Fontes, J. V. et al., 2020. Patterns and vertical loads in water shipping in systematic wet dam-break experiments. *Ocean Engineering,* Volume 197.

Hirotada, H., Grenier, N., Sueyoshi, M. & Le Touzé, D., 2022. Comparison of MPS and SPH methods for solving forced motion ship flooding problems. *Applied Ocean Research,* Volume 118.

Huang, L. et al., 2022. A review on the modelling of wave-structure interactions based on OpenFOAM. *OpenFOAM Journal,* Volume 2, pp. 116-142.

Jandaghian, M. & Shakibaeinia, A., 2020. An enhanced weakly-compressible MPS method for free-surface flows. *Computer Methods in Applied Mechanics and Engineering,* Volume 360.

Kashani, A. H., Halabian, A. M. & Asghari, K., 2018. A numerical study of tuned liquid damper based on incompressible SPH method combined with TMD analogy. *Journal of Fluids and Structures,* October, Volume 82, p. 394–411.

Khayyer, A. & Gotoh, H., 2013. Enhancement of performance and stability of MPS mesh-free particle method for multiphase flows characterized by high density ratios. *Journal of Computational Physics,* Volume 242, pp. 211-233.

Koshizuka, S. & Oka, Y., 1996. Moving-particle semi-implicit method for fragmentation of incompressible fluid. *Nuclear Science and Engineering,* 123(3), pp. 421-434.

Lee, B.-H., Park, J.-C., Kim, M.-H. & Hwang, S.-C., 2011. Step-by-step improvement of MPS method in simulating violent free-surface motions and impact-loads. *Computer Methods in Applied Mechanics and Engineering,* 200(9-12), pp. 1113-1125.

Lee, C. S. et al., 2012. Investigation of structural responses of breakwaters for green water based on fluid-structure interaction analysis. *International Journal of Naval Architecture and Ocean Engineering,* 4(2), pp. 83-95.

Lo, E. Y. M. & Shao, S., 2002. Simulation of near-shore solitary wave mechanics by an incompressible SPH method. *Applied Ocean Research,* 24(5), pp. 275-286.

Lucy, L., 1977. A numerical approach to the testing of the fission hypothesis. *Astronomical Journal,* Volume 82, pp. 1013-1024.

Mazhar, F. et al., 2021. On the meshfree particle methods for fluid-structure interaction problems. *Engineering Analysis with Boundary Elements,* Volume 124, pp. 14-40.

McNamara, K. P., Awad, B. N., Tait, M. J. & Love, J. S., 2021. Incompressible smoothed particle hydrodynamics model of a rectangular tuned liquid damper containing screens. *Journal of Fluids and Structures*, Volume 103.

Roache, P., 1998. *Verification and Validation in Computational Science and Engineering.* Albuquerque: Hermosa Publishers.

Roache, P. J., 2009. *Fundamentals of Verification and Validation.* s.l.:Hermosa Publishers.

Rosetti, G. F. et al., 2019. CFD and experimental assessment of green water events on an FPSO Hull section in beam waves. *Marine Structures,* Volume 65, pp. 154-180.

Shakibaeinia, A. & Jin, Y.-C., 2010. A weakly compressible MPS method for modeling of open-boundary free-surface flow. *International journal for numerical methods in fluids,* 63(10), pp. 1208-1232.

Shibata, K., Koshizuka, S., Matsunaga, T. & Massaie, I., 2017. The overlapping particle technique for multi-resolution simulation of particle methods. *Computer Methods in Applied Mechanics and Engineering,* Volume 325, pp. 434-462.

Silva, D. F. C. & Rossi, R. R., 2014. Green water loads determination for fpso exposed to beam sea conditions. *33rd International Conference on Ocean, Offshore and Arctic Engineering.*

Sun, Y. et al., 2023. A conservative particle splitting and merging technique with dynamic pattern and minimum density error. *Engineering Analysis with Boundary Elements,* May, Volume 150, p. 246–258.

Tanaka, M., Cardoso, R. & Bahai, H., 2018. Multi-resolution MPS method. *Journal of Computational Physics*, Volume 359, pp. 106-136.

Taylor, K. E., 2001. Summarizing multiple aspects of model performance in a single diagram. *Journal of Geophysical Research: Atmospheres,* Volume 106, pp. 7183-7192.

Temarel, P. et al., 2016. Prediction of wave-induced loads on ships: Progress and challenges. *Ocean Engineering*, Volume 119, pp. 274-308.

Tsukamoto, M. M., Cheng, L.-Y. & Motezuki, F. K., 2016. Fluid interface detection technique based on neighborhood particles centroid deviation (NPCD) for particle methods. *International Journal for Numerical Methods in Fluids*, 82(3), pp. 148-168.

Tsukamoto, M. M., Cheng, L. Y. & Nishimoto, K., 2011. Analytical and numerical study of the effects of an elastically-linked body on sloshing. *Computers & Fluids,* 49(1), pp. 1-21.

Vacondio, R. et al., 2020. Grand challenges for Smoothed Particle Hydrodynamics numerical schemes. *Computational Particle Mechanics,* September, Volume 8, p. 575–588.

Wang, L., Jiang, Q. & Zhang, C., 2017. Improvement of moving particle semi-implicit method for simulation of progressive water waves. *International Journal for Numerical Methods in Fluids,* 85(2), pp. 69-89.

Weber, L. J., Cherian, M. P., Allen, M. E. & Muste, M., 2000. *Headloss characteristics for perforated plates and flat bar screens: Technical Report 411,* s.l.: Iowa Institute of Hydraulic Engineering.

Wei, Z. et al., 2018. Chaos in breaking waves. *Coastal Engineering,* Volume 140, pp. 272-291.